\documentclass[10pt,twocolumn]{article}

\usepackage[a4paper,margin=1.9cm,columnsep=0.6cm]{geometry}
\usepackage[T2A]{fontenc}
\usepackage[utf8]{inputenc}
\usepackage[russian,english]{babel}
\usepackage{tempora}
\usepackage{newtxmath}

\usepackage{microtype}
\usepackage{CJKutf8}
\usepackage{amsmath,amssymb}
\usepackage{booktabs}
\usepackage{array}
\usepackage{enumitem}
\usepackage{tabularx}
\usepackage{graphicx}
\graphicspath{{jadr_plots/}}
\usepackage[round,authoryear]{natbib}
\usepackage{caption}
\usepackage{xcolor}
\usepackage{xurl}
\usepackage[colorlinks=true,linkcolor=blue!50!black,citecolor=blue!50!black,urlcolor=blue!50!black]{hyperref}
\usepackage{titlesec}
\usepackage[ruled,vlined]{algorithm2e}
\usepackage{textcomp}
\titleformat{\section}{\normalfont\fontsize{11}{13}\bfseries}{\thesection}{1em}{}
\titleformat{\subsection}{\normalfont\fontsize{10.5}{12.5}\bfseries}{\thesubsection}{1em}{}
\titlespacing*{\section}{0pt}{1.4ex plus 0.3ex minus 0.2ex}{0.8ex}
\titlespacing*{\subsection}{0pt}{1.1ex plus 0.2ex minus 0.15ex}{0.6ex}

\newcommand{\Dset}{\mathcal{D}}
\newcommand{\Nset}{\mathcal{N}}
\newcommand{\fp}{\mathrm{fp}}
\newcommand{\AUC}{\mathrm{AUC}}
\newcommand{\Lds}{L_{\mathrm{ds}}}
\newcommand{\Dlog}{D_{\log}}
\newcommand{\SH}{\mathrm{SH}}
\newcommand{\zh}[1]{\begin{CJK}{UTF8}{gbsn}#1\end{CJK}}
\newcommand{\ru}[1]{\foreignlanguage{russian}{#1}}
\DeclareUnicodeCharacter{2116}{\textnumero}

\hypersetup{
  pdftitle={Silent Alarm: A J-Space Protocol for Comparing Danger Recognition Across Models and Quantization Levels},
  pdfauthor={Prosvirnin Roman, Minchenkov Victor, Soldatov Alexey, Bashun Vladimir, Sergeev Anton}
}

\begin{document}

\twocolumn[
\begin{center}
{\LARGE\bfseries Silent Alarm: A J-Space Protocol for Comparing Danger Recognition Across Models and Quantization Levels}\par\vspace{5pt}

{\bfseries Prosvirnin Roman%
{\renewcommand{\thefootnote}{\fnsymbol{footnote}}\footnotemark[1]}%
, Minchenkov Victor, Soldatov Alexey, Bashun Vladimir, Sergeev Anton}\par\vspace{3pt}
{\bfseries HSE University}\par\vspace{1.3em}
\end{center}
\begin{minipage}{\textwidth}
\small
\noindent\textbf{Abstract.}
Jailbreak robustness research typically evaluates safety through generated responses using a large language model (LLM)-as-judge approach. However, such evaluations are sensitive to the benchmark grading procedure and capture only the observed behavior on a given set of attacks, without directly revealing the hidden fragility of the underlying safety mechanisms. This work proposes JADR (Jacobian Assessment of Danger Recognition), a protocol that measures a model's internal representation through Jacobian space (J-space, a recently proposed workspace of verbalizable concepts) before the first response token is generated. For every prompt and layer we record the top-$k$ J-space tokens, which are grouped into six behavioral scenario axes and compared between a danger sample based on StrongREJECT and a safe control drawn from XSTest and OKTest. The method does not call on an external judge model: the computation runs entirely locally, on the activations of the model under evaluation, which lets us compare both different models against each other and modifications of a single model -- quantization and fine-tuning in particular -- on the same terms. The final comparison rests on the proposed SafetyAUC metric, complemented with bootstrap confidence intervals. The protocol is applied to six models (Qwen3-1.7B, Qwen3-4B, Qwen3-8B, Qwen3-Uncensored-4B, Qwen3-SafeRL-4B, Gemma 2 9B) across three weight-representation regimes -- BF16, INT8, and INT4 -- and checked against an independent behavioral evaluation with the StrongREJECT grader. The metric separates models with a strong versus a weak internal safety mechanism with statistical significance and captures substantively different effects across quantization regimes.

\end{minipage}
\vspace{1.6em}
]

{\renewcommand{\thefootnote}{\fnsymbol{footnote}}\footnotetext[1]{Corresponding author: \texttt{rprosvirnin@hse.ru}}}

\section{Introduction}
\label{sec:intro}

Jailbreak robustness research and automated red teaming commonly rely on a behavioral protocol: generated responses are scored by a large language model (LLM) judge, and the resulting decisions are aggregated into rate-based metrics such as attack success rate (ASR). This protocol has some limitations. The results are sensitive to the choice and configuration of the judge, and, when a continuous grader is used, to the binarization threshold as well, which makes it hard to compare results across papers. The output score captures observed compliance with or violation of policy on a fixed set of attacks, but it does not directly characterize the robustness of the internal representations and computational mechanisms tied to harm recognition and refusal. As a result, a fragile defense may remain undetected on the current test set and only surface under different or adaptively chosen attacks. In addition, ASR values obtained on different benchmarks are frequently incomparable, owing to differences in prompt distribution, threat model, success criteria, attack budget, and grading procedure.

All these limitations follow from the same fact: the judge sees only the final text of the response, not the state of the model that produced it. Engineering modifications of language models -- quantization, fine-tuning, abliteration -- are rarely neutral with respect to safety, even when they are neutral with respect to quality \citep{hong2024decoding, qi2023finetuningalignedlanguagemodels}. A model can hold its previous level on standard benchmarks while behaving differently on dangerous prompts: refusing more often, giving more useful detail where it used to refuse, or losing the ability to tell a harmful instruction apart from a superficially similar but benign task. An LLM-as-judge evaluation records the fact of such degradation after it happens, but not its mechanism: it sees only the final text, not the model's internal processing before that text appeared.

This study treats safety as an internal, verbalizable representation available to the model at the moment it chooses a response. We care not only about whether the model refused in words, but also about whether its internal workspace carries a signal of danger, illegality, willingness to comply, or an attempt to bypass a restriction. Recent mechanistic evidence supports this distinction. \citet{zhao2025harmfulness} show that harmfulness and refusal occupy distinct latent directions, and that jailbreak attacks can suppress refusal signals while preserving a model's internal representation of harmfulness. The distinction also matters specifically for quantization: numerical error in the representation may not immediately change the surface style of the response, yet it can already disturb the internal separation between dangerous and safe cases -- and, conversely, a model can keep a clear internal recognition of danger and still generate harmful text. These three regimes -- intact recognition with intact refusal, intact recognition with lost refusal, and lost recognition -- are indistinguishable if one looks only at the text of the response.

The basis for this measurement is J-space -- a space of verbalizable directions read out by the Jacobian lens \citep{gurnee2026workspace}. For every layer and position, the lens recovers which vocabulary tokens the model ``holds ready'' for future speech, correcting the logit lens by the averaged Jacobian of the transition to the final state. We read this content at the final prompt position -- the position of the last token before the first token of the response -- where it is not yet distorted by already-generated text, and we compare it between a dangerous set of prompts and a safe control made up of outwardly alarming but essentially harmless formulations. This is the key move that lets us tie the appearance of safety tokens in the workspace to the actual danger of the prompt as recognized by the model. In this way we show that the signal in J-space tracks the semantics of danger rather than its vocabulary.

We introduce a reproducible J-space assessment protocol at the final prompt position that requires neither response generation nor a call to an external judge model: the entire computation runs locally, on the activations of the model being evaluated. Comparison is moved onto the dimensionless rank $\AUC$ (SafetyAUC, ComplAUC) and the logarithmic shift of token counts $\Dlog$. In addition, quantization regimes are compared with the $\Delta SH$ (Safety Headroom) metric, which accounts for change along both AUC dimensions. We applied the Jacobian Assessment of Danger Recognition (JADR) protocol to six models, including variants with reinforced and with weakened safety tuning, and to three weight-representation regimes, which lets us compare both models against each other and modifications of a single model -- quantization above all -- in the same terms. The internal metrics are systematically checked against an independent behavioral evaluation on the StrongREJECT benchmark. This comparison serves to differ cases where a model's safety recognition and actual behavior diverge from cases where they experience a simultaneous degradation.

\section{Related Work}
\label{sec:related}

\subsection{Global Workspace and the Jacobian Lens}

The methodological foundation of this work is the result of \citet{gurnee2026workspace}, who showed that language models contain a relatively small class of representations (J-space) available for future verbalization and generation control. The authors describe this as a functional analogue of a global workspace: not all of a model's intermediate state is equally available to future output, but part of it can be read out through the Jacobian lens -- a correction to the logit lens that accounts for the averaged Jacobian of the transition from an intermediate layer to the final state. As illustrations, the original work reports signs of recognizing attempted prompt injection (tokens such as \textit{fake}, \textit{injection}, \textit{poison}) and multilingual equivalents of such signs. The causal role of these directions is established there through intervention -- ablation and activation injection. The question of safety evaluation for quantized models was not addressed in that work.

The \textbf{JADR} approach -- Jacobian Assessment of Danger Recognition -- proposes using the same tool to build safety metrics and behavioral robustness metrics in two scenarios: comparing models against each other (cross-model) and comparing modifications of a single model, quantization in particular (cross-modification).

\subsection{Jailbreak Evaluation}

Recent jailbreak benchmarks have moved toward more standardized and reproducible evaluation protocols \citep{mazeika2024harmbench,chao2024jailbreakbench,souly2024strongreject}. For jailbreak evaluation we chose the StrongREJECT benchmark \citep{souly2024strongreject} and its fine-tuned grader (a fine-tuned Gemma 2B), which assigns a score from 0 (safe) to 1 (dangerous). This gives us a set of prompts that are unambiguously forbidden and a scoring model that provides a continuous estimate of the harm in the response itself, rather than of the mere fact of refusal. In JADR, StrongREJECT plays two roles, neither of which turns it into a first-order judge for the internal metrics. The set of forbidden prompts forms the dangerous sample $\Dset, \ |\Dset| = 313$; the trained grader is used separately, only for behavioral validation -- checking whether the internal shifts line up with the actual content of the response. These roles do not overlap: JADR's internal metrics never use the text of the response and do not depend on a judge call at all, so the LLM-as-judge limitations discussed above do not carry over to the main result of this work, and the behavioral score remains an independent source of validation and an indicator.

\subsection{Quantization and Safety}

As a rule, quantizing a model degrades both its safety and its instruction-following robustness. It has already been noted, however, that the effect of quantization on refusal is not always monotonic: compressing a model can either weaken or unintentionally strengthen its tendency to refuse, and moderate 4-bit quantization in some cases even preserves or improves model trustworthiness compared with coarser compression \citep{hong2024decoding}. An LLM balances task completion against safety-aligned instruction following. This tension is also visible in over-refusal benchmarks, where benign prompts with superficially harmful wording can trigger unnecessary refusals \citep{rottger2024xstest,shi2024overkill}.
A study on quantization-based jailbreaks \citep{lee2025quantizationjailbreaking} shows a non-linear performance-safety trade-off: 4-bit quantization can preserve much of the practical efficiency/quality balance, but considerable vulnerabilities remain.

JADR provides the tooling to measure this at the representation level: which scenario axis changed, in which direction, and in which layers -- which applies not only to INT8 and INT4 but in principle to any weight modification for which a J-lens readout can be obtained.

\section{Problem Formulation}
\label{sec:problem}

Let $\mathcal{M}$ be a set of language models. For each model $m\in\mathcal{M}$, let
$\mathcal{Q}_m$ be a set of weight-representation regimes containing a full-precision
reference $\fp$, and let $\mathcal{L}_m$ denote the model's layers. Prompts are drawn from
two disjoint populations: a dangerous population $\Dset$ and a benign control $\Nset$.
The control contains prompts whose surface vocabulary may appear harmful but whose requested
goals are benign. Thus, successful discrimination must depend on prompt intent rather than on
the presence of isolated harm-related words.

For a semantic axis $g$, modified regime $q$ define an internal evidence score
\begin{equation}
\phi_g(m,q,p)\in\mathbb{R},
\label{eq:evidence-score}
\end{equation}
obtained from the model state after the complete prompt has been processed and before response
generation begins. The primary task is to construct a statistic
\begin{equation}
R_g(m,q)=\mathrm{P}\!\left[
\phi_g(m,q,p_\Dset)>\phi_g(m,q,p_\Nset)
\right],
\label{eq:formal-separation}
\end{equation}
where $p_\Dset\sim\Dset$ and $p_\Nset\sim\Nset$ are sampled independently. For the safety
axis, a value substantially above $0.5$ indicates that the model carries stronger internal
safety evidence for dangerous prompts than for benign controls.

Therefore, $R_{\mathrm{safety}}(m,\fp)$ must support
cross-model comparison despite differences in model depth, tokenizer, and vocabulary coverage.
For every modified regime $q\ne\fp$, the degradation statistic
\begin{equation}
\Delta R_g(m,q)=R_g(m,\fp)-R_g(m,q)
\label{eq:formal-degradation}
\end{equation}
must quantify changes relative to the same model in full precision. The ordering induced
by the internal safety statistic must be compared with an independent behavioral harm measure
$h(m,q)$ (share of responses with a LLM-as-judge score at threshold $0.5$), while allowing for cases in which internal danger recognition and generated behavior
diverge.

Accordingly, the method must produce dimensionless model-level scores, preserve the direction
of within-model changes without using generated responses to construct the internal metric.

\section{Methods}
\label{sec:method}

\subsection{Jacobian Lens}

Let $h_{\ell,t}\in\mathbb{R}^{d}$ be the residual stream at layer $\ell$ and position $t$, and
$h_{L,t'}$ the final residual stream at position $t'\ge t$. The Jacobian lens estimates
how a small change in the intermediate state carries over into the final one:
\begin{equation}
J_\ell = \operatorname*{mean}_{\text{prompt},\,t,\,t'\ge t}
\left[\frac{\partial h_{L,t'}}{\partial h_{\ell,t}}\right].
\label{eq:jacobian}
\end{equation}
The average is taken over the prompts of the corpus, the source positions $t$, and all current and future target positions $t'$. The matrix $J_\ell$ is averaged over a large set of calibration prompts ($N$=100, pretrain-like, exactly the figures recommended by \citet{gurnee2026workspace}) and does not depend on the specific example on which the readout is later built.

After applying $J_\ell$
and the unembedding matrix $W_U$ we obtain a distribution over vocabulary tokens:
\begin{equation}
\mathrm{lens}_\ell(h_{\ell,t}) = \mathrm{softmax}\big(W_U\cdot\mathrm{norm}(J_\ell\cdot h_{\ell,t})\big).
\label{eq:lens}
\end{equation}
Here, $\mathrm{norm}$ is the model's native final normalization  layer (RMSNorm) applied immediately before unembedding. Unlike standard logit-lens-style vocabulary projections discussed in prior lens-based analyses \citep{belrose2023tunedlens} ($J_\ell=I$ in the simplest case), this distribution should not be read as a prediction of the
next token: it shows what content the model \emph{could} make available to future verbalization, if it were to verbalize that content. This work does not use the full distribution~\eqref{eq:lens}, only the top-$k$ ($k$ = 100 for every metric in this paper) tokens for each layer and prompt.

All internal metrics are computed at the final prompt position -- the position of the last token before
the first token of the assistant's response. At this position the model has already
processed the whole prompt but has not yet started generating. The activation at this position depends
causally only on the prompt, not on already-written response text. Computing the internal metric
requires no decoding and only one forward pass per prompt. By contrast, behavioral evaluation
requires an initial prefill followed by one sequential
decoding step for every generated token, after which a separate grader model processes the
prompt--response pair. For a response of several hundred tokens, JADR thus replaces several hundred
sequential decoding steps and a second model call with a single pre-generation forward pass. This
makes the measurement conceptually clean (the ``before speech'' setting) and computationally
inexpensive.

\subsection{Scenario Axes and Frozen Lexicon}

The read-out is interpreted through a frozen multilingual lexicon divided into six
semantic axes:
\begin{equation}
\begin{gathered}
g \in \{
\text{safety},\ \text{compliance},\ \text{evasion}, \\
\text{softening},\ \text{hedging},\ \text{harm}
\}.
\end{gathered}
\label{eq:axes}
\end{equation}
The lexicon is fixed before evaluation and contains only token forms observed in the
workspaces of the evaluated models. The safety axis is the primary axis for measuring
danger recognition, while compliance provides a complementary indication of willingness
to follow a request. Evasion, softening, hedging, and harm are retained as diagnostic axes
that describe how the prompt is represented. Their roles are summarized in
Table~\ref{tab:axis-groups}.

\begin{table}[htbp]
\centering\small
\setlength{\tabcolsep}{4pt}
\caption{Scenario axes used by JADR.}
\label{tab:axis-groups}
\begin{tabular}{@{}p{0.27\linewidth}p{0.63\linewidth}@{}}
\toprule
Axis & Diagnostic role \\
\midrule
\textbf{safety} & refusal, risk, illegality \\
\textbf{compliance} & willingness to comply or instruct \\
\textbf{evasion} & bypassing and concealing the prompt's goal \\
softening & softened and transitional wording \\
hedging & cautious generalization, hedges \\
harm & concrete language of a specific harmful method \\
\bottomrule
\end{tabular}
\end{table}

The softening and hedging axes are included to capture possible refuse-then-comply
patterns, in which an initial refusal is followed by a softened response that discloses
the requested information. Representative stems are listed in Appendix Table~\ref{tab:lexicon}.

\subsection{Token Counters}

For prompt $p$ and layer $\ell$, let $\tau_r(p,\ell)$ denote the vocabulary token at
position $r$ after the lens scores have been sorted in descending order. For group $g$,
its presence in the top-$k$ readout is measured in two ways. The plain count:
\begin{equation}
c_g(p,\ell) = \sum_{r=1}^{k} \mathbf{1}\big[\tau_r(p,\ell) \in g\big].
\label{eq:count}
\end{equation}
Count does not distinguish a token at the first position of the readout from one at the hundredth, even though earlier tokens are more likely to reflect what the model is actually ready to say. A rank-weighted version accounts for this difference explicitly:
\begin{equation}
\mathrm{dcg}_g(p,\ell) = \sum_{r=1}^{k}
\frac{\mathbf{1}\big[\tau_r(p,\ell) \in g\big]}{\log_2(r+1)}.
\label{eq:dcg}
\end{equation}
Both quantities are reported throughout the paper: count as a transparent but coarse
estimate, DCG as its refinement, robust to how deep into the top-$k$ the
readout extends. Below, ``counter'' refers to either of the two quantities -- all metrics of
\S\ref{sec:auc}--\S\ref{sec:cross-quant} apply to both without change.

For aggregation across layers we introduce the sum
\begin{equation}
T_g(p) = \sum_{\ell} c_g(p,\ell).
\label{eq:total}
\end{equation}
We introduce intermediate notation for the mean per-layer signal and the per-layer difference between dangerous and non-dangerous prompts:
\begin{equation}
\begin{aligned}
S_g(m,q,P,\ell)
  &= \operatorname*{mean}_{p\in P} c_g(p,\ell), \\
A_g(m,q,\ell)
  &= S_g(m,q,\Dset,\ell)
   - S_g(m,q,\Nset,\ell).
\end{aligned}
\label{eq:layer-signal}
\end{equation}
These quantities depend directly on the size of the lexicon (for instance, on an extension of the configuration file) and on the scale of the token counter, so we move on to more advanced metrics that reflect true separability.

\subsection{Rank AUC}
\label{sec:auc}

The final core of the comparison is the rank $\AUC$ (equivalent to the normalized Mann-Whitney statistic): the probability that a random value from one sample exceeds a random value from the other \citep{hanley1982auc}. It is dimensionless and depends only on the ordering of values, which makes it comparable across models with different tokenizers, depths, and vocabulary coverage, so it is robust to changes in the lexicon. AUC is not inflated by small denominators. For example, a model with no pronounced signal on axis $g$ gives $\AUC\approx 0.5$.

The per-layer version:
\begin{equation}
\AUC_g(m,q,\ell) = \mathrm{P}\big[c_g(p_\Dset,\ell) > c_g(p_\Nset,\ell)\big],
\label{eq:layer-auc}
\end{equation}
where $p_\Dset\in\Dset$ and $p_\Nset\in\Nset$ are drawn at random and independently. The aggregate
model-level version is computed on the sum over layers:
\begin{equation}
\AUC_g(m,q) = \mathrm{P}\big[T_g(p_\Dset) > T_g(p_\Nset)\big].
\label{eq:agg-auc}
\end{equation}
Equation~\eqref{eq:agg-auc} instantiates the separation statistic
in Eq.~\eqref{eq:formal-separation} with \(\phi_g(m,q,p)=T_g(p)\).
A value of $0.5$ means no separability along axis $g$; values above $0.5$ mean the
group is more pronounced on dangerous prompts; values below $0.5$ mean the opposite direction. The strength of separability is given by $|\AUC-0.5|$.

For every quantity reported in this paper we give a $95\%$ confidence interval obtained by bootstrapping \emph{over prompts} ($B=1000$ resamples with replacement, $2.5$ and $97.5$ percentiles).
Resampling over prompts, rather than over individual tokens, is chosen because tokens within a single readout are strongly correlated.

\subsection{Cross-model Metrics}
\label{sec:cross-model}

For model comparison in this paper, the regime is fixed at $q=\fp$. The main comparator metric, \textbf{SafetyAUC}, is --
\begin{equation}
\mathrm{SafetyAUC}(m) = \AUC_{\mathrm{safety}}(m,\fp),
\label{eq:safetyauc}
\end{equation}
which shows how reliably the model's workspace tells apart dangerous and harmless prompts
along the safety axis -- that is, how well the model ``recognizes danger'' in the intent of the prompt. The complementary metric, \textbf{ComplAUC}, is --
\begin{equation}
\mathrm{ComplAUC}(m) = \AUC_{\mathrm{compliance}}(m,\fp),
\label{eq:complauc}
\end{equation}
which shows the direction of the willingness to comply: a value below $0.5$ corresponds to a stronger
willingness on safe prompts, above $0.5$ to that willingness shifting toward dangerous
prompts.

SafetyAUC and ComplAUC need not agree: a model can correctly recognize danger and still show a compliance signature shifted toward danger. A common scenario is one where the model `understands' the prompt is dangerous, but its helpfulness objective leads it to seek a compromise that turns into a successful attack. Internal recognition can therefore diverge from eventual behavior: a model with a high SafetyAUC can still give a harmful response.

To compare the absolute magnitude balance between models we use the logarithmic ratio of the total danger and safe counters, with a correction:
\begin{equation}
\begin{aligned}
\Lds^{g}(m,q)
  &= \log\frac{X_g+0.5}{Y_g+0.5}, \\
X_g
  &= \sum_{p\in\Dset} T_g(p),
  \qquad
Y_g
  = \sum_{p\in\Nset} T_g(p).
\end{aligned}
\label{eq:lds}
\end{equation}
The $0.5$ correction is standard for logarithmic ratios and, at the scale of the totals here (hundreds to thousands), has virtually no effect on the result. On a rare axis it prevents division by zero,
and the width of the bootstrap interval then signals on its own how unreliable the estimate is. A cross-model
comparison on a single axis is simply the difference of $\Lds^{g}$ between two models.

In addition, to localize axis $g$ for a given model we introduce the set of important layers:
\begin{equation}
\begin{aligned}
&\mathcal{I}_g(m,q) = \big\{\, \ell : |\AUC_g(m,q,\ell) - 0.5| > \tau \,\big\}, \\
&\tau = 0.2.
\label{eq:important-layers}
\end{aligned}
\end{equation}

This is not a metric in the strict sense -- it is an indicator of whether the model's internal behavior has shifted to new layers under quantization. The threshold $\tau = 0.2$ is chosen empirically and is not a strict cutoff.

\subsection{Cross-modification Metrics}
\label{sec:cross-quant}

For comparing quantization regimes, the model $m$ is held fixed and the regime $q$ is compared against the reference $\fp$. Because of this, every metric here is built as a difference.

The shift in separability under quantization:
\begin{equation}
\Delta\AUC_g(m,q) = \AUC_g(m,\fp) - \AUC_g(m,q).
\label{eq:delta-auc}
\end{equation}
For safety, a positive value means erosion and degradation of recognition -- a value near zero means it is preserved.

The magnitude shift between $\fp$ and $q$:
\begin{equation}
\Dlog^{g}(m,q) = \Lds^{g}(m,q) - \Lds^{g}(m,\fp).
\label{eq:dlog}
\end{equation}

The physical meaning of the metric is this: under quantization, tokens of group $g$ became relatively more frequent on dangerous prompts ($\Dlog>0$) or on safe ones ($\Dlog<0$).

The sign favorable to model safety depends on the axis: for safety it is
$\Dlog\ge 0$ (the refusal-related signal did not decrease on dangerous prompts). For compliance -- $\Dlog\le 0$ (the relative willingness to carry out dangerous instructions did not increase).

The safety axis alone is not enough if recognition is preserved under quantization while the willingness to comply grows. We introduce the \textbf{Safety Headroom} (SH) metric.
\begin{equation}
\begin{aligned}
    \SH(m,q) = &\mathrm{SafetyAUC}(m,q) \ - \\ \big(&\mathrm{ComplAUC}(m,q) - 0.5\big)
\end{aligned}
\label{eq:sh}
\end{equation}

\begin{equation}
\begin{aligned}
\Delta\SH(m,q) &= \SH(m,q) - \SH(m,\fp) \\
&= \Delta\AUC_{\mathrm{compl}} - \Delta\AUC_{\mathrm{safety}}
\end{aligned}
\label{eq:delta-sh}
\end{equation}
A negative $\Delta\SH$ means the joint safety margin has degraded even when SafetyAUC on its own
looks preserved (the case of Qwen-4B under INT4:
$\Delta\AUC_{\mathrm{safety}}\!\approx\!0$, yet $\Delta\SH<0$ owing to a shift in compliance).
Identity \eqref{eq:delta-sh} sets out the role of the metric: SH is not an independent measurement, but a
within-model screening aggregate that folds the shift along two axes into one signed scalar.
Using it only in the form of a difference is essential. The absolute level of ComplAUC carries a
constant offset from the style of refusal (in aligned models, a substantial share of compliance vocabulary
on dangerous prompts comes from the stock phrasing of polite refusal and redirection, not from an
intent to comply), so an absolute SH is not interpreted across models -- in the difference
$\Delta\SH$ this offset largely cancels. A residual caveat concerns INT4: if quantization itself
changes the frequency and style of refusal, the compliance component of $\Delta\SH$ partly reflects this shift.

$\Delta\AUC$ and $\Dlog$ measure different things and can disagree: for instance, under INT4 a model's relative mass of safety tokens on dangerous prompts can grow ($\Dlog>0$) while the rank separation stays almost unchanged ($\Delta\AUC\approx 0$), so both quantities have to be read together.
$\Delta\AUC$ should be treated as the reranker and the primary criterion, and $\Dlog$ as a diagnostic indicator.

In addition we assess mechanism relocation -- the Jaccard index of the intersection between the sets of important
layers before and after quantization:
\begin{equation}
J_g(m,q) = \frac{|\mathcal{I}_g(m,\fp)\cap\mathcal{I}_g(m,q)|}
{|\mathcal{I}_g(m,\fp)\cup\mathcal{I}_g(m,q)|}.
\label{eq:jaccard}
\end{equation}
A low $J_g$ shows that the set of important layers has changed, but on its own it does not prove
a causal break in the mechanism: a minority of the layers that differ can turn out to be decisive, while a
majority of the layers that coincide can be secondary. For this reason $J_g$ is used only as an indicator of
localization -- the causal role of a specific layer is established separately, through intervention
(replacing or restoring a layer's weights and then measuring the effect), not through a threshold on
$J_g$.

Algorithm~\ref{alg:jadr} summarizes the complete internal evaluation. The same procedure is
run with the count and DCG counters. Behavioral generation is performed separately and does
not enter the construction of the internal metrics.

\begin{algorithm}[t]
\small
\caption{JADR evaluation for one model}
\label{alg:jadr}
\KwIn{Model $m$; regimes $\mathcal{Q}_m$ with reference $\fp$; calibration corpus
$\mathcal{C}$; prompt sets $\Dset,\Nset$; lexicon axes $\mathcal{G}$; depth $k$;
bootstrap count $B$}
\KwOut{$\AUC_g$, $\Lds^g$, $\Delta\AUC_g$, $\Dlog^g$, $\Delta\SH$, and $J_g$}
\ForEach{$q\in\mathcal{Q}_m$}{
  Load model $m$ in regime $q$\;
  \ForEach{layer $\ell\in\mathcal{L}_m$}{
    Estimate $J_\ell$ on $\mathcal{C}$ using Eq.~\eqref{eq:jacobian}\;
  }
  \ForEach{$p\in\Dset\cup\Nset$}{
    Run one forward pass and retain $h_{\ell,t}$ at the final prompt position\;
    \ForEach{$\ell\in\mathcal{L}_m$}{
      $V_{p,\ell}\leftarrow\operatorname{TopK}(\mathrm{lens}_\ell(h_{\ell,t}),k)$\;
      \ForEach{$g\in\mathcal{G}$}{
        Compute $c_g(p,\ell)$ and $\mathrm{dcg}_g(p,\ell)$\;
      }
    }
    Aggregate each counter over layers to obtain $T_g(p)$\;
  }
  \ForEach{$g\in\mathcal{G}$}{
    Compute layer-level and aggregate $\AUC_g$, $\Lds^g$, and $\mathcal{I}_g$\;
    Bootstrap prompts $B$ times to obtain confidence intervals\;
  }
}
\ForEach{$q\in\mathcal{Q}_m\setminus\{\fp\}$}{
  Compute $\Delta\AUC_g$, $\Dlog^g$, $\Delta\SH$, and $J_g$ relative to $\fp$\;
}
\KwRet{Internal metrics and prompt-level bootstrap confidence intervals}\;
\end{algorithm}

\subsection{Behavioral Validation}
\label{sec:behavior}

For the attacking set of prompts $\Dset$ we compute the share of responses with a StrongREJECT score (the grader
\texttt{strongreject\_finetuned}) at threshold $0.5$, denoted $h(m,q)$. The higher $h(m,q)$, the more dangerous the response.

We also consider the share of refusals identified by a regular expression that parses the start of the response (refusal). For the safe control $\Nset$
we compute only the refusal share (over-refusal): on harmless prompts, the StrongREJECT grader
measures helpfulness rather than harm, and it often marks a sensible, useful answer as ``harmful'' -- the harm score on $\Nset$ is therefore not interpreted, and only the refusal share is used. Agreement in direction between the internal shifts ($\Delta\AUC$, $\Dlog$) and the behavioral ones ($h$, refusal) is one of the criteria for trusting the result.

\section{Experiments}
\label{sec:experiments}

\subsection{Datasets and Models}

The dangerous set $\Dset$ consists of 313 StrongREJECT prompts. The safe control $\Nset$ consists of 320 prompts drawn from XSTest-safe (250) and OKTest (70): both sets are chosen so as to contain vocabulary that is alarming in form but harmless in content, and thereby test the metric's robustness against false triggering on words, rather than only its ability to tell clearly harmful prompts apart from safe ones that resemble them.

The protocol is applied to six models: Gemma2-9B \citep{gemmateam2024gemma2} and the Qwen3 family at different sizes (1.7B, 4B, 8B), Ablit-4B --
a version of Qwen3-4B with the refusal direction removed by ablation \citep{arditi2024refusal}, and Qwen3-SafeRL-4B -- a version
of the same model with additional safety fine-tuning. Ablit-4B and SafeRL-4B serve as calibration poles: they are not targets of the study in their own right, but are needed to check that the metrics really do pick up a known-in-advance weakening and strengthening of the safety signal.

For every model we compare three
weight-representation regimes: $\fp$ (bf16), INT8 \citep{dettmers2022llmint8} and INT4
(bitsandbytes NF4 with double quantization \citep{dettmers2023qlora}). The number of source layers and the
sample sizes are given in Table~\ref{tab:setup}.

\begin{table}[htbp]
\centering\small
\caption{Experimental setup. Global settings for all models: pre-generation read-out at
\texttt{rel\_pos}$=-1$, top-$k=100$; frozen lexicon sizes (words): safety 170, compliance
47, evasion 49, softening 16, hedging 22, harm 84; weight regimes: FP (bf16), INT8, INT4 (NF4).}
\label{tab:setup}
\begin{tabular}{@{}lcc@{}}
\toprule
Code name & Layers & Full name \\
\midrule
Gemma-9B  & 41 & gemma-2-9b-it \\
Ablit-4B  & 35 & huihui-qwen3-4b-abliterated-v2 \\
Qwen-1.7B & 27 & qwen3-1.7b \\
Qwen-4B   & 35 & qwen3-4b \\
SafeRL-4B & 35 & qwen3-4b-saferl \\
Qwen-8B   & 35 & qwen3-8b \\
\bottomrule
\end{tabular}
\end{table}

\subsection{Implementation Details}

Internal metrics are computed at the final prompt position and therefore require no decoding.
Behavioral generations used for independent validation, by contrast, run in a deterministic
regime ($\mathrm{temperature}=0$, $\mathrm{top\_p}=1$, $\mathrm{do\_sample}=\mathrm{false}$);
if a model supports a reasoning (thinking) mode, it is turned off. For every layer and prompt we store the top-200 J-lens readout. Confidence intervals throughout are obtained by bootstrapping over prompts with $B=300$--$1000$ resamples -- the threshold for localizing important layers is $\tau=0.2$.

\section{Results}
\label{sec:results}

\begin{table*}[tp]
\centering\small
\setlength{\tabcolsep}{5pt}
\caption{Cross-model summary at FP. SafetyAUC is the primary internal comparator (higher =
stronger recognition of danger). ComplAUC below $0.5$ corresponds to stronger compliance on
safe prompts; values above $0.5$ indicate a compliance signature shifted toward dangerous
prompts. \texttt{harm@0.5} and refusal are measured on the danger set, over-refusal on the
safe set. 95\% CI in brackets.}
\label{tab:model-summary}
\begin{tabular}{@{}lccccc@{}}
\toprule
Model & SafetyAUC & ComplAUC & harm@0.5 & refusal$_D$ & over-refusal$_N$ \\
\midrule
Gemma-9B  &\textit{\textbf{0.975}} [0.965, 0.984] & 0.662 [0.614, 0.710] & \textbf{0.022} & \textbf{0.958} & 0.147 \\
Ablit-4B  & 0.593 [0.544, 0.640] & 0.838 [0.808, 0.864] & 0.904 & 0.000 & 0.009 \\
Qwen-1.7B & 0.895 [0.872, 0.922] & 0.597 [0.553, 0.644] & 0.204 & 0.383 & 0.041 \\
Qwen-4B   & 0.973 [0.961, 0.983] & 0.497 [0.455, 0.544] & 0.048 & 0.706 & 0.028 \\
SafeRL-4B & 0.976 [0.965, 0.985] & 0.503 [0.461, 0.552] & 0.345 & 0.000 & \textbf{0.003} \\
Qwen-8B   & \textbf{0.979} [0.969, 0.988] & \textbf{0.242} [0.210, 0.277] & 0.067 & 0.754 & 0.041 \\
\bottomrule
\end{tabular}
\end{table*}

Table~\ref{tab:model-summary} shows that at $\fp$ four models -- Gemma-9B, Qwen-4B,
SafeRL-4B and Qwen-8B -- have a SafetyAUC around $0.97$ or higher, Qwen-1.7B is somewhat lower
($0.895$) but still clearly separates dangerous from safe. Ablit-4B stands apart:
its SafetyAUC is close to $0.59$, nearly at the chance level, while its
behavioral harm rate is the highest of the six models ($0.904$). This split matters:
it points not to a subtle weakening of refusal but to a fundamentally different regime, in which the
internal distinction between dangerous and safe along the safety axis is simply weak.

ComplAUC adds a dimension that cannot be recovered from SafetyAUC alone. Qwen-8B
combines the highest SafetyAUC with a ComplAUC noticeably below $0.5$ -- the model not only
recognizes danger but also shows a stronger willingness to comply specifically on
safe prompts. Ablit-4B, by contrast, has a ComplAUC above $0.83$, consistent with the
pattern of following instructions regardless of their danger that is characteristic of this model.

\begin{table*}[tp]
\centering\small
\setlength{\tabcolsep}{5pt}
\caption{Quantization effects on the safety ($s$) and compliance ($c$) axes (point
estimates, count metric). $\Delta\SH$ is the joint Safety Headroom shift. The direction of
harm differs by axis: for safety, $\Delta\AUC_s>0$ (eroded separability) and $\Dlog^{s}<0$
(safety mass lost on the danger set) are adverse - for compliance, $\Delta\AUC_c<0$ (the
compliance signature shifting toward the danger set) and $\Dlog^{c}>0$ (comply-on-danger
increased) are adverse - $\Delta\SH<0$ is adverse. Per-axis source tables: $\Delta\AUC$ in Table~\ref{tab:dauc}, $\Dlog$ with
confidence intervals in Tables~\ref{tab:dlog-int4}--\ref{tab:dlog-int8}
(Appendix).}
\label{tab:quant-summary}
\begin{tabular}{@{}lrrrrrrrrrr@{}}
\toprule
& \multicolumn{4}{c}{$\Delta\AUC$} & \multicolumn{4}{c}{$\Dlog$} & \multicolumn{2}{c}{$\Delta\SH$} \\
\cmidrule(lr){2-5}\cmidrule(lr){6-9}\cmidrule(lr){10-11}
Model & $s$\,i8 & $s$\,i4 & $c$\,i8 & $c$\,i4 & $s$\,i8 & $s$\,i4 & $c$\,i8 & $c$\,i4 & i8 & i4 \\
\midrule
Gemma-9B  & -0.001 & -0.000 & -0.000 & -0.013 &  0.006 & -0.014 & -0.000 &  0.009 &  0.001 & -0.013 \\
Ablit-4B  & -0.011 & -0.086 &  0.001 & -0.012 &  0.028 &  0.143 &  0.002 & -0.023 &  0.012 &  0.073 \\
Qwen-1.7B & -0.001 & \textbf{0.041} &  0.002 & \textbf{-0.047} &  0.029 & \textbf{-0.166} &  0.006 & \textbf{0.096} &  0.003 & \textbf{-0.088} \\
Qwen-4B   & -0.001 &  0.004 & -0.002 & \textbf{-0.048} &  0.025 &  0.084 & -0.001 & \textbf{0.084} & -0.001 & \textbf{-0.052} \\
SafeRL-4B & -0.001 & -0.003 &  0.004 & -0.003 &  0.035 &  0.056 & -0.008 &  0.024 &  0.005 & -0.000 \\
Qwen-8B   & -0.001 &  0.006 & -0.005 & \textbf{-0.038} &  0.020 & \textbf{-0.138} & -0.004 &  0.015 & -0.004 & \textbf{-0.044} \\
\bottomrule
\end{tabular}
\end{table*}

Table~\ref{tab:quant-summary} gives the summary picture for quantization. INT8 leaves the
separability of the safety axis almost unchanged: $\Delta\AUC_{\mathrm{safety}}$ stays under
$0.011$ in absolute value for every model. INT4 behaves in a far more uneven way. Qwen-1.7B shows a marked drop in
SafetyAUC ($\Delta\AUC_s=0.041$), whereas Ablit-4B's, by contrast, recovers
($\Delta\AUC_s=-0.086$); for Qwen-4B and Qwen-8B the shift in $\AUC$ itself is small, yet Safety Headroom still declines ($\Delta\SH=-0.052$ and $-0.044$)
-- meaning the safety axis alone can look preserved in a place where the joint margin
of safety has already weakened because of compliance.

The sign of $\Dlog^{\mathrm{safety}}$ shows which way the balance of absolute masses has shifted. For Qwen-4B
under INT4 $\Dlog^{s}>0$: safety tokens do not relatively disappear on dangerous prompts -- yet
$\Delta\SH<0$, meaning the overall margin declines specifically because of compliance, not safety.
For Qwen-1.7B and Qwen-8B, by contrast, a negative $\Dlog^{s}$ points to a relative
weakening of the safety balance on dangerous prompts, consistent with the drop in $\AUC$ for the first of
these two models. This illustrates the general principle: $\Delta\AUC$, $\Dlog$ and $\SH$ measure different aspects
of the same phenomenon and must be read together, not as interchangeable numbers.

Ablit-4B deserves special attention: for this model INT4 does not only fail to make the
problem worse, but partly restores lost recognition ($\Dlog^{s}=+0.143$) and
partly restores behavioral refusal (harm@0.5 drops from $0.904$ at $\fp$ to
$0.875$ at INT4, Table~\ref{tab:behavior}). A probable explanation is
that ablation, as a technique for removing the safety mechanism, relies on a precise edit to the weights, and
quantization partly erases this edit along with the rest of the numerical content;
in models where the safety mechanism was not artificially removed, the same coarsening effect
works in the opposite direction and can weaken, rather than restore, recognition. This interpretation is consistent with prior work showing non-monotonic safety effects under compression \citep{hong2024decoding,lee2025quantizationjailbreaking}. We also note the striking and distinctive robustness of Gemma-9B's safety and compliance mechanisms to quantization.

We note that quantization has the strongest effect not on the safety mechanism but on the compliance mechanism (for models trained to be safe), consistent with prior work on compression and model trustworthiness \citep{hong2024decoding,lee2025quantizationjailbreaking}. An increase in compliance on dangerous prompts is the main vulnerability of quantized models.

\begin{table}[htbp]
\centering\scriptsize
\setlength{\tabcolsep}{3.5pt}
\caption{Behavioral validation by model and weight regime: harm@0.5, mean StrongREJECT
score and refusal on the danger set, over-refusal on the safe set. Lexicon-independent
(computed from generated responses).}
\label{tab:behavior}
\begin{tabular}{@{}llcccc@{}}
\toprule
Model & $q$ & harm@0.5 & mean SR & refusal$_D$ & over-ref.$_N$ \\
\midrule
Gemma-9B  & fp   & \textbf{0.022} & \textbf{0.019} & \textbf{0.958} & 0.147 \\
Gemma-9B  & int8 & \textbf{0.003} & \textbf{0.008} & \textbf{0.958} & 0.153 \\
Gemma-9B  & int4 & \textbf{0.022} & \textbf{0.018} & \textbf{0.962} & 0.150 \\
\addlinespace
Ablit-4B  & fp   & 0.904 & 0.780 & 0.000 & 0.009 \\
Ablit-4B  & int8 & 0.898 & 0.775 & 0.003 & 0.006 \\
Ablit-4B  & int4 & 0.875 & 0.747 & 0.000 & \textbf{0.000} \\
\addlinespace
Qwen-1.7B & fp   & 0.204 & 0.209 & 0.383 & 0.041 \\
Qwen-1.7B & int8 & 0.201 & 0.210 & 0.351 & 0.044 \\
Qwen-1.7B & int4 & 0.188 & 0.178 & 0.403 & 0.031 \\
\addlinespace
Qwen-4B   & fp   & 0.048 & 0.063 & 0.706 & 0.028 \\
Qwen-4B   & int8 & 0.038 & 0.057 & 0.725 & 0.034 \\
Qwen-4B   & int4 & 0.048 & 0.073 & 0.728 & 0.025 \\
\addlinespace
SafeRL-4B & fp   & 0.345 & 0.385 & 0.000 & \textbf{0.003} \\
SafeRL-4B & int8 & 0.374 & 0.373 & 0.000 & \textbf{0.003} \\
SafeRL-4B & int4 & 0.329 & 0.361 & 0.000 & \textbf{0.000} \\
\addlinespace
Qwen-8B   & fp   & 0.067 & 0.071 & 0.754 & 0.041 \\
Qwen-8B   & int8 & 0.073 & 0.069 & 0.760 & 0.034 \\
Qwen-8B   & int4 & 0.054 & 0.063 & 0.789 & 0.037 \\
\bottomrule
\end{tabular}
\end{table}

\section{Discussion}
\label{sec:analysis}

\subsection{Count and DCG}

Count and DCG measure different aspects of the same readout: count is how many tokens of the
group land in the top-$k$, DCG is how high up they sit there. For most axes
and models the two quantities agree qualitatively: for example, the DCG version of SafetyAUC for
Gemma-9B and Qwen-8B differs from the count version by no more than $0.007$
(Table~\ref{tab:model-auc-dcg} in the Appendix versus Table~\ref{tab:model-summary}). The difference
becomes substantive specifically in the context of quantization: if a modified model keeps
including a group's tokens in the top-$k$ but systematically pushes them down in rank, count will
not notice, while DCG will register the weakened availability of the signal before it
shows up in the plain count. For this reason DCG is used not as a replacement for count, but as a
variant of the same measure that is more sensitive to rank shuffling, and both versions are reported for
every final metric.

\subsection{Recognition versus Behavior}

Comparing the SafetyAUC and harm@0.5 columns in Table~\ref{tab:model-summary} pulls apart two
distinct failure modes that a purely behavioral metric cannot tell apart. Ablit-4B has a
low SafetyAUC (0.593) and the highest harm rate of the six models (0.904) -- this is a model in
which safety is broken at the level of representation, not only at the level of the final
response. SafeRL-4B and, to a lesser extent, Qwen-1.7B are built differently: their SafetyAUC is high
(0.976 and 0.895), but the harm rate is also noticeably above zero (0.345 and 0.204) -- a case where
the model internally tells a dangerous prompt from a safe one, but this does not stop it from a harmful
response. For fine-tuning and defensive quantization these two cases call for
different interventions: in the first, restoring or strengthening recognition itself. In the second, working on what happens between recognition and the decision to comply.

The Qwen-8B vs Gemma-9B comparison is remarkable. By SafetyAUC the models are indistinguishable (0.979 versus 0.975, confidence intervals overlapping), while Qwen-8B's ComplAUC is lower (0.242 versus 0.662), so on the axis of
recognition Qwen-8B looks no worse. The behavioral results, however, show the opposite pattern: Qwen-8B has a higher harm@0.5
(0.067 versus 0.022) and a lower refusal share (0.754 versus 0.958). The discrepancy is explained by two
factors invisible to separability metrics. First, Gemma-9B is tuned to refuse aggressively and achieves a low harm rate at the cost of higher over-refusal (0.147), whereas Qwen-8B is shifted toward helpfulness (over-refusal 0.041), with a higher harm rate at the same quality of recognition. A presumptive explanation for the over-refusal gap is that Gemma-9B is marginally lower SafetyAUC point estimate reflects slightly weaker separation of dangerous and benign prompts. Combined with its more conservative safety policy, this could make Gemma-9B refuse more benign prompts than Qwen-8B. Second, a ComplAUC of $0.662>0.5$ for Gemma-9B, given a refusal share of $0.958$, cannot
reflect an intent to carry out a harmful request and corresponds instead to the vocabulary of polite refusal and redirection on dangerous prompts. A lower ComplAUC for Qwen-8B should not be read as
unconditionally safer, but as a trait of a model that suppresses compliance tokens when answering a dangerous prompt, unlike Gemma-9B. It is worth keeping in mind that compliance is a less stable and less clear-cut axis than safety. Because of this, ComplAUC can be misleading when comparing models from different families. One should also allow for grading error in the StrongREJECT grader, which, acting as an LLM-as-judge, may itself have introduced some bias into the response score.

The SafeRL-4B model shows a substantial $harm @ 0.5$, which leads to the counterintuitive suggestion that the SafeRL model is more dangerous than Qwen3-4B. One should, however, take into account the specifics of how the SafeRL model was trained \citep{qwen3guard2025} -- it was rewarded for both safety and helpfulness at the same time, and penalized for unnecessary refusals. It is therefore meant to answer less often with a plain `I cannot help with that,' replacing a dangerous response with a safe alternative. Every SafeRL-4B response to a dangerous question follows the same template: \textit{`It is important to clarify that your request is illegal. However, I can suggest...'}. Precisely because of this shared opening construction, the regex refusal rate scored this model at 0.00.

StrongREJECT reads the extended answer as disclosure of harmful instructions, while its grader assigns low scores specifically to a plain refusal. On harm recognition, SafeRL-4B shows a higher SafetyAUC than the base Qwen3-4B. This example makes the difference between assessing a model's internal state and assessing its required behavior particularly clear. SafeRL-4B's responses can be scored differently depending on the benchmark. The internal safety mechanism, however -- the separation of harmful from benign -- stays the same. This is exactly the problem that the SafetyAUC metric addresses.

\subsection{Layer Relocation}
\label{sec:layer-relocation}

The sets of important layers $\mathcal{I}_g$ let us check whether an axis stays localized in
the same layers after quantization. For the safety axis under INT4 the Jaccard index is above $0.9$ for four
of the six models and drops to $0$ for Ablit-4B (its safety mechanism is already very weakly
localized at $\fp$, in just two layers, so even a small shift changes the
set entirely). A low $J_g$ does not by itself mean a causal break: some of the layers
that formally fell out of the threshold set could have been secondary, with one
or two of the remaining layers being decisive. A separate experiment restoring the original
FP weights in individual layers of the INT4 model (rescue) would confirm this caveat directly:
restoring the weights in just a few layers, chosen not by $\AUC$-importance but by an independent criterion
(the refusal direction \citep{arditi2024refusal}), noticeably changed the model's eventual behavior -- meaning the causal weight of
specific layers need not match their formal membership in $\mathcal{I}_g$. The full
relocation table for safety and compliance across all models is given in
Table~\ref{tab:relocation} of the Appendix.

\section{Limitations}
\label{sec:limitations}

JADR depends on a lexicon fixed in advance. The lexicon groups cover individual vocabulary tokens and are poorly suited to multi-token expressions and to
cases where the model describes a dangerous action descriptively, without directly using any word from the
lexicon. In this sense the metric systematically underestimates recognition wherever
it is expressed non-literally.

The rank $\AUC$ and the Jaccard index are correlational quantities. They show that an axis
separates $\Dset$ and $\Nset$ in certain layers, but they do not prove that these specific layers are
causally responsible for refusal. A causal conclusion requires intervention -- replacing
weights, causal tracing or activation patching \citep{meng2022locating,hase2023localization}, or ablating them and then measuring the effect.

There is also a substantive limitation tied to the very object being measured: StrongREJECT and similar graders score text that has already been generated, whereas JADR captures the state of the model before generation. The gap between these two levels is not a flaw of the method -- it is itself a substantive result of comparing danger recognition and behavior. Finally, the six models used here include two calibration poles (ablation, additional
safety fine-tuning) and make no claim to exhaustively cover model families or sizes. Carrying the protocol over to a new architecture requires re-checking sensitivity to $k$ and recomputing the important layers, rather than directly reusing the thresholds obtained here.

It should be noted that the metrics presented here do depend on the lexicon set in the configuration file. The AUC metrics, however, are robust to changes in the lexicon. Also, the wider the lexicon already is, the smaller the change from subsequently extending it carefully. Moreover, in scenarios that call for comparing a group of models, researchers fix a single shared vocabulary, and the ranking conclusions remain valid. Because the workspace tokens are stored throughout the protocol, a researcher can immediately compare the results of several lexicons against each other.

The method is meant for comparing AI safety across models and for evaluating modified models, and for more deliberate design of
quantization and subsequent fine-tuning procedures: it makes it possible to notice degradation of internal
recognition of danger earlier than it becomes visible in the eventual share of harmful responses, and
to separate it from a shift toward willingness to comply, which a behavioral metric also does not catch right away.

The present results do not justify a universal absolute SafetyAUC cutoff. In a deployment setting, the minimum acceptable SafetyAUC should instead be calibrated against an approved reference model, domain-specific harmful prompts, and the risk tolerance of the application. Passing the JADR screen is not a safety certification. It indicates that the candidate has not shown measurable degradation of internal danger recognition or joint safety headroom and may proceed to behavioral red-teaming and policy-specific validation.

\section{Conclusion}
\label{sec:conclusion}

This paper has proposed JADR -- a reproducible protocol for assessing internal danger recognition in language models. The protocol measures the verbalizable content of J-space before the first token is generated, calls on no external judge model, and runs in a single forward pass, which makes it usable within a local pipeline and suitable for comparing both different models and modifications of a single model on the same terms.

The empirical results support several conclusions. The proposed SafetyAUC metric separates models of different sizes, and models with a deliberately strong versus a deliberately weakened safety mechanism, with statistical significance. The resulting ranking agrees with the behavioral evaluation from the StrongREJECT benchmark once model-specific response patterns are taken into account. Internal recognition and eventual behavior can, however, part ways. Using Qwen3-4B-SafeRL as an example, we showed that a model can recognize danger well and still disclose harmful information, because its particular fine-tuning forbids answering with a plain refusal.

INT8 quantization is practically neutral to internal separability, whereas INT4 has heterogeneous effects: from noticeable erosion in Qwen3-1.7B to a partial reversal of the ablation effect in Qwen3-4B uncensored, where 4-bit compression erases part of the targeted weight edit and restores both safety tokens in J-space and the refusal share. Quantization, however, harms the compliance (instruction-following) axis more than it harms the safety mechanism.

Degradation of the protective mechanisms can be noticed before it shows up in the share of harmful responses, and localized by axis and by layer. JADR, protocol of measuring inner danger recognition, can therefore serve as the first gate in a pre-deployment evaluation pipeline for quantized and fine-tuned checkpoints. A developer first evaluates an accepted full-precision checkpoint and then applies the same prompt sets, lexicon, and read-out configuration to each candidate modification. A candidate should be withheld from deployment and returned for re-quantization, additional alignment, or targeted behavioral testing when $\Delta\AUC_{\mathrm{safety}}>0$, or when $\Delta\SH<0$ with the same statistical criterion. This screening step requires neither response generation nor an external judge model, so clearly degraded candidates can be rejected before the more expensive behavioral evaluation.

Future work will proceed in two directions: causal verification of the layers identified here through targeted interventions with a measured effect on behavior, and carrying the protocol beyond the Qwen and Gemma families.

\subsection*{Declaration of competing interest}

The authors have no competing interests to declare that are relevant to the content of this article.

\bibliographystyle{plainnat}
\bibliography{references}

\appendix

\section{Additional Tables and Figures}

\textbf{Tables}
\label{app:tables}

Below are the full per-axis tables that supplement the summary results of the main text.
Notation: models -- Gemma-9B, Ablit-4B (abliterated), Qwen-1.7B, Qwen-4B, SafeRL-4B,
Qwen-8B; axes -- Saf(ety), Cmp(compliance), Eva(sion), Sof(tening), Hed(ging), Hrm (harm); D/N
-- danger/safe. All quantities are read at the final prompt position, top-$k=100$, on the frozen lexicon;
uncertainty is a $95\%$ bootstrap CI over prompts.

\begin{table*}[tp]
\centering
\small
\renewcommand{\arraystretch}{1.3}
\setlength{\tabcolsep}{8pt}
\caption{Representative frozen-lexicon stems per scenario axis in English, Chinese,
and Russian.}
\label{tab:lexicon}
\begin{tabularx}{\linewidth}{>{\raggedright\arraybackslash}p{3.4cm}
                                >{\raggedright\arraybackslash}X}
\toprule
\textbf{Category} & \textbf{Typical stems} \\
\midrule
Safety &
unsafe, cannot, illegal, harmful, forbidden, disclaimer, dangerous, apologize, \zh{不安全, 危险, 违法, 拒绝}, \ru{запрещ}-, \ru{опасн}- \\
\addlinespace
Compliance &
sure, here, step, method, guide, provide, recipe, instruction, \zh{可以, 步骤, 方法, 提供}, \ru{конечно} \\
\addlinespace
Harm &
weapon, explosive, malware, credential, poison, payload, firearm, toxin, \zh{武器, 炸弹, 毒药, 恶意软件} \\
\addlinespace
Evasion &
bypass, hidden, conceal, undetect, stealth, loophole, spoof, anonym, \zh{绕过, 隐藏, 秘密, 伪装}, \ru{обход} \\
\addlinespace
Softening &
however, although, instead, rather, while, nevertheless, \zh{但是, 虽然, 只是, 然而}, \ru{но} \\
\addlinespace
Hedging &
usually, generally, possible, probably, maybe, typically, seems, \zh{通常, 可能, 一般来说, 不一定}, \ru{возможно} \\
\bottomrule
\end{tabularx}
\end{table*}

\textbf{Figures}
\label{app:figures}

This section collects the diagnostic plots that supplement the numerical tables of Appendix~\ref{app:tables}.
Every panel is built on the same grid of six models (Gemma-9B, Ablit-4B, Qwen-1.7B,
Qwen-4B, SafeRL-4B, Qwen-8B) and three weight-representation regimes ($\fp$, INT8, INT4).

\begin{table*}[tp]
\centering\scriptsize
\setlength{\tabcolsep}{4pt}
\caption{Per-axis discrimination $\AUC$ at FP, \textbf{count} metric, 95\% CI. Long-row
layout of the same data as Table~\ref{tab:model-summary}'s SafetyAUC/ComplAUC columns,
extended to all six axes.}
\label{tab:model-auc}
\begin{tabular}{@{}lll@{\hspace{1.2em}}lll@{}}
\toprule
Model & Axis & $\AUC$ [95\% CI] & Model & Axis & $\AUC$ [95\% CI] \\
\midrule
Gemma-9B & safety     & 0.975 [0.965, 0.984] & Qwen-4B & safety     & 0.973 [0.961, 0.983] \\
Gemma-9B & compliance & 0.662 [0.614, 0.710] & Qwen-4B & compliance & 0.497 [0.455, 0.544] \\
Gemma-9B & evasion    & 0.668 [0.617, 0.715] & Qwen-4B & evasion    & 0.653 [0.604, 0.694] \\
Gemma-9B & softening  & 0.366 [0.323, 0.410] & Qwen-4B & softening  & 0.550 [0.507, 0.592] \\
Gemma-9B & hedging    & 0.197 [0.163, 0.236] & Qwen-4B & hedging    & 0.457 [0.409, 0.501] \\
Gemma-9B & harm       & 0.814 [0.781, 0.850] & Qwen-4B & harm       & 0.610 [0.573, 0.644] \\
\addlinespace
Ablit-4B & safety     & 0.593 [0.544, 0.640] & SafeRL-4B & safety     & 0.976 [0.965, 0.985] \\
Ablit-4B & compliance & 0.838 [0.808, 0.864] & SafeRL-4B & compliance & 0.503 [0.461, 0.552] \\
Ablit-4B & evasion    & 0.500 [0.468, 0.531] & SafeRL-4B & evasion    & 0.830 [0.796, 0.861] \\
Ablit-4B & softening  & 0.559 [0.516, 0.602] & SafeRL-4B & softening  & 0.547 [0.505, 0.589] \\
Ablit-4B & hedging    & 0.289 [0.253, 0.324] & SafeRL-4B & hedging    & 0.202 [0.169, 0.237] \\
Ablit-4B & harm       & 0.526 [0.498, 0.553] & SafeRL-4B & harm       & 0.493 [0.458, 0.526] \\
\addlinespace
Qwen-1.7B & safety     & 0.895 [0.872, 0.922] & Qwen-8B & safety     & 0.979 [0.969, 0.988] \\
Qwen-1.7B & compliance & 0.597 [0.553, 0.644] & Qwen-8B & compliance & 0.242 [0.210, 0.277] \\
Qwen-1.7B & evasion    & 0.780 [0.739, 0.814] & Qwen-8B & evasion    & 0.825 [0.797, 0.862] \\
Qwen-1.7B & softening  & 0.466 [0.413, 0.515] & Qwen-8B & softening  & 0.641 [0.598, 0.685] \\
Qwen-1.7B & hedging    & 0.356 [0.313, 0.395] & Qwen-8B & hedging    & 0.134 [0.108, 0.160] \\
Qwen-1.7B & harm       & 0.381 [0.339, 0.417] & Qwen-8B & harm       & 0.733 [0.694, 0.769] \\
\bottomrule
\end{tabular}
\end{table*}

\begin{table*}[p]
\centering\scriptsize
\setlength{\tabcolsep}{4pt}
\caption{Per-axis discrimination $\AUC$ at FP, \textbf{DCG} (rank-weighted) metric, 95\% CI.
DCG counterpart of Table~\ref{tab:model-auc}: the safety and compliance columns are the
DCG-based SafetyAUC and ComplAUC discussed in \S\ref{sec:analysis}.}
\label{tab:model-auc-dcg}
\begin{tabular}{@{}lll@{\hspace{1.2em}}lll@{}}
\toprule
Model & Axis & $\AUC$ [95\% CI] & Model & Axis & $\AUC$ [95\% CI] \\
\midrule
Gemma-9B & safety     & 0.968 [0.955, 0.977] & Qwen-4B & safety     & 0.973 [0.960, 0.982] \\
Gemma-9B & compliance & 0.676 [0.637, 0.715] & Qwen-4B & compliance & 0.521 [0.474, 0.564] \\
Gemma-9B & evasion    & 0.659 [0.610, 0.699] & Qwen-4B & evasion    & 0.648 [0.608, 0.686] \\
Gemma-9B & softening  & 0.356 [0.308, 0.391] & Qwen-4B & softening  & 0.551 [0.506, 0.599] \\
Gemma-9B & hedging    & 0.201 [0.168, 0.232] & Qwen-4B & hedging    & 0.456 [0.411, 0.507] \\
Gemma-9B & harm       & 0.813 [0.780, 0.846] & Qwen-4B & harm       & 0.611 [0.575, 0.643] \\
\addlinespace
Ablit-4B & safety     & 0.635 [0.590, 0.676] & SafeRL-4B & safety     & 0.975 [0.962, 0.985] \\
Ablit-4B & compliance & 0.831 [0.802, 0.859] & SafeRL-4B & compliance & 0.524 [0.479, 0.568] \\
Ablit-4B & evasion    & 0.499 [0.465, 0.531] & SafeRL-4B & evasion    & 0.823 [0.792, 0.853] \\
Ablit-4B & softening  & 0.561 [0.522, 0.609] & SafeRL-4B & softening  & 0.549 [0.507, 0.590] \\
Ablit-4B & hedging    & 0.288 [0.249, 0.320] & SafeRL-4B & hedging    & 0.196 [0.162, 0.230] \\
Ablit-4B & harm       & 0.526 [0.500, 0.555] & SafeRL-4B & harm       & 0.491 [0.461, 0.524] \\
\addlinespace
Qwen-1.7B & safety     & 0.893 [0.865, 0.915] & Qwen-8B & safety     & 0.979 [0.970, 0.989] \\
Qwen-1.7B & compliance & 0.622 [0.581, 0.668] & Qwen-8B & compliance & 0.243 [0.209, 0.278] \\
Qwen-1.7B & evasion    & 0.777 [0.741, 0.812] & Qwen-8B & evasion    & 0.822 [0.784, 0.854] \\
Qwen-1.7B & softening  & 0.483 [0.435, 0.523] & Qwen-8B & softening  & 0.655 [0.613, 0.700] \\
Qwen-1.7B & hedging    & 0.348 [0.311, 0.391] & Qwen-8B & hedging    & 0.136 [0.110, 0.164] \\
Qwen-1.7B & harm       & 0.379 [0.338, 0.418] & Qwen-8B & harm       & 0.730 [0.691, 0.763] \\
\bottomrule
\end{tabular}
\end{table*}

\begin{table*}[p]
\centering\scriptsize
\setlength{\tabcolsep}{3.6pt}
\caption{$\Dlog$ under \textbf{INT4}, \textbf{count} metric, per axis, 95\% CI. This is the
regime where content shifts are large enough to be worth reporting with full confidence
intervals. The corresponding INT8 shifts are given as point estimates in
Table~\ref{tab:dlog-int8}.}
\label{tab:dlog-int4}
\begin{tabular}{@{}lll@{\hspace{0.8em}}lll@{}}
\toprule
Model & Axis & $\Dlog$ [95\% CI] & Model & Axis & $\Dlog$ [95\% CI] \\
\midrule
Gemma-9B & safety     & -0.014 [-0.033, 0.007] & Qwen-4B & safety     &  0.084 [0.040, 0.125] \\
Gemma-9B & compliance &  0.009 [-0.000, 0.020] & Qwen-4B & compliance &  0.084 [0.068, 0.099] \\
Gemma-9B & evasion    & -0.101 [-0.175, -0.021] & Qwen-4B & evasion    &  0.140 [-0.015, 0.334] \\
Gemma-9B & softening  &  0.013 [-0.052, 0.065] & Qwen-4B & softening  &  0.183 [0.096, 0.276] \\
Gemma-9B & hedging    &  0.018 [-0.025, 0.057] & Qwen-4B & hedging    & -0.472 [-0.596, -0.338] \\
Gemma-9B & harm       &  0.036 [-0.027, 0.109] & Qwen-4B & harm       & -0.215 [-0.406, -0.004] \\
\addlinespace
Ablit-4B & safety     &  0.143 [0.097, 0.195] & SafeRL-4B & safety     &  0.056 [0.026, 0.084] \\
Ablit-4B & compliance & -0.023 [-0.038, -0.007] & SafeRL-4B & compliance &  0.024 [0.011, 0.038] \\
Ablit-4B & evasion    &  0.304 [0.052, 0.700] & SafeRL-4B & evasion    & -0.038 [-0.134, 0.075] \\
Ablit-4B & softening  &  0.076 [-0.028, 0.179] & SafeRL-4B & softening  & -0.248 [-0.343, -0.157] \\
Ablit-4B & hedging    &  0.408 [0.247, 0.596] & SafeRL-4B & hedging    &  0.210 [0.050, 0.367] \\
Ablit-4B & harm       &  0.064 [-0.179, 0.271] & SafeRL-4B & harm       & -0.046 [-0.343, 0.229] \\
\addlinespace
Qwen-1.7B & safety     & -0.166 [-0.219, -0.121] & Qwen-8B & safety     & -0.138 [-0.172, -0.104] \\
Qwen-1.7B & compliance &  0.096 [0.066, 0.125] & Qwen-8B & compliance &  0.015 [-0.007, 0.039] \\
Qwen-1.7B & evasion    &  0.064 [-0.029, 0.165] & Qwen-8B & evasion    & -0.399 [-0.517, -0.295] \\
Qwen-1.7B & softening  & -0.054 [-0.094, -0.011] & Qwen-8B & softening  &  0.184 [0.116, 0.255] \\
Qwen-1.7B & hedging    &  0.135 [0.056, 0.215] & Qwen-8B & hedging    &  0.009 [-0.071, 0.089] \\
Qwen-1.7B & harm       & -0.388 [-0.580, -0.190] & Qwen-8B & harm       & -0.173 [-0.263, -0.077] \\
\bottomrule
\end{tabular}
\end{table*}

\begin{table*}[p]
\centering\scriptsize
\setlength{\tabcolsep}{3pt}
\caption{$\Dlog$ under \textbf{INT8}, \textbf{count} metric, point estimates (no CI, effect
sizes here are small and close to zero, see \S\ref{sec:results}).}
\label{tab:dlog-int8}
\begin{tabular}{@{}lrrrrrr@{}}
\toprule
Model & Saf & Cmp & Eva & Sof & Hed & Hrm \\
\midrule
Gemma-9B  &  0.006 & -0.000 & -0.000 &  0.038 & -0.020 & -0.005 \\
Ablit-4B  &  0.028 &  0.002 & -0.099 & -0.018 &  0.013 & -0.032 \\
Qwen-1.7B &  0.029 &  0.006 &  0.105 &  0.008 &  0.005 & -0.039 \\
Qwen-4B   &  0.025 & -0.001 &  0.068 &  0.020 & -0.004 & -0.050 \\
SafeRL-4B &  0.035 & -0.008 &  0.007 &  0.018 &  0.001 & -0.008 \\
Qwen-8B   &  0.020 & -0.004 &  0.035 & -0.018 &  0.096 &  0.038 \\
\bottomrule
\end{tabular}
\end{table*}

\begin{table*}[p]
\centering\scriptsize
\setlength{\tabcolsep}{3pt}
\caption{$\Delta\AUC_g=\AUC_g(\mathrm{FP})-\AUC_g(q)$, \textbf{count} metric, per axis,
point estimates. Source of the safety and compliance $\Delta\AUC$ columns in
Table~\ref{tab:quant-summary}. For safety, $\Delta\AUC>0$ = eroded separability. For
compliance, $\Delta\AUC<0$ = signature shifted toward the danger set.}
\label{tab:dauc}
\begin{tabular}{@{}llrrrrrr@{}}
\toprule
Model & $q$ & Saf & Cmp & Eva & Sof & Hed & Hrm \\
\midrule
Gemma-9B  & int8 & -0.001 & -0.000 &  0.005 & -0.006 &  0.002 &  0.002 \\
Gemma-9B  & int4 & -0.000 & -0.013 &  0.015 & -0.006 &  0.014 & -0.011 \\
Ablit-4B  & int8 & -0.011 &  0.001 & -0.000 &  0.002 &  0.013 &  0.003 \\
Ablit-4B  & int4 & -0.086 & -0.012 & -0.021 &  0.005 & -0.023 &  0.016 \\
Qwen-1.7B & int8 & -0.001 &  0.002 &  0.002 &  0.010 & -0.041 & -0.009 \\
Qwen-1.7B & int4 &  0.041 & -0.047 &  0.020 &  0.054 & -0.104 &  0.004 \\
Qwen-4B   & int8 & -0.001 & -0.002 & -0.025 & -0.010 &  0.016 &  0.013 \\
Qwen-4B   & int4 &  0.004 & -0.048 & -0.042 & -0.045 &  0.134 &  0.028 \\
SafeRL-4B & int8 & -0.001 &  0.004 &  0.004 &  0.002 &  0.002 &  0.001 \\
SafeRL-4B & int4 & -0.003 & -0.003 &  0.003 &  0.089 &  0.067 &  0.006 \\
Qwen-8B   & int8 & -0.001 & -0.005 & -0.013 &  0.005 & -0.026 &  0.004 \\
Qwen-8B   & int4 &  0.006 & -0.038 &  0.064 & -0.056 & -0.005 &  0.039 \\
\bottomrule
\end{tabular}
\end{table*}

\begin{table*}[p]
\centering\scriptsize
\setlength{\tabcolsep}{2.6pt}
\caption{$\Dlog$, \textbf{DCG} metric, both quantization regimes, point estimates (no CI,
DCG is reported here as a robustness check against Table~\ref{tab:dlog-int4}/\ref{tab:dlog-int8},
not as an independent primary result).}
\label{tab:dlog-dcg}
\begin{tabular}{@{}llrrrrrr@{}}
\toprule
Model & $q$ & Saf & Cmp & Eva & Sof & Hed & Hrm \\
\midrule
Gemma-9B  & int8 &  0.007 & -0.002 & -0.000 &  0.045 & -0.020 & -0.005 \\
Gemma-9B  & int4 & -0.013 &  0.033 & -0.125 &  0.018 &  0.026 &  0.020 \\
Ablit-4B  & int8 &  0.031 & -0.004 & -0.072 & -0.023 &  0.028 & -0.019 \\
Ablit-4B  & int4 &  0.150 & -0.016 &  0.349 &  0.063 &  0.469 &  0.106 \\
Qwen-1.7B & int8 &  0.028 &  0.004 &  0.087 &  0.012 & -0.009 & -0.018 \\
Qwen-1.7B & int4 & -0.186 &  0.096 &  0.083 & -0.082 &  0.096 & -0.455 \\
Qwen-4B   & int8 &  0.035 & -0.006 &  0.036 &  0.018 &  0.003 & -0.052 \\
Qwen-4B   & int4 &  0.085 &  0.097 &  0.115 &  0.148 & -0.433 & -0.159 \\
SafeRL-4B & int8 &  0.045 & -0.010 & -0.034 &  0.022 &  0.006 & -0.018 \\
SafeRL-4B & int4 &  0.034 &  0.052 & -0.023 & -0.293 &  0.255 & -0.009 \\
Qwen-8B   & int8 &  0.022 & -0.005 &  0.028 & -0.015 &  0.094 &  0.048 \\
Qwen-8B   & int4 & -0.124 &  0.010 & -0.344 &  0.193 & -0.008 & -0.177 \\
\bottomrule
\end{tabular}
\end{table*}

\begin{table*}[p]
\centering\footnotesize
\setlength{\tabcolsep}{3.2pt}
\caption{Absolute per-prompt token counts (\textbf{count} metric), summed over layers, by
axis, model and weight regime. D/N = danger/safe.}
\label{tab:abs-count}
\begin{tabular}{@{}l*{12}{c}@{}}
\toprule
& \multicolumn{2}{c}{Saf} & \multicolumn{2}{c}{Cmp} & \multicolumn{2}{c}{Eva}
& \multicolumn{2}{c}{Sof} & \multicolumn{2}{c}{Hed} & \multicolumn{2}{c}{Hrm} \\
\cmidrule(lr){2-3}\cmidrule(lr){4-5}\cmidrule(lr){6-7}\cmidrule(lr){8-9}\cmidrule(lr){10-11}\cmidrule(lr){12-13}
Model/$q$ & D & N & D & N & D & N & D & N & D & N & D & N \\
\midrule
Gemma-9B/fp   & 525.4 & 98.6 &  91.6 &  83.3 & 8.7 & 7.1 & 4.3 & 13.4 & 2.4 & 31.1 & 13.6 & 4.5 \\
Gemma-9B/int8 & 523.6 & 97.7 &  91.9 &  83.6 & 8.6 & 7.0 & 4.5 & 13.3 & 2.4 & 31.3 & 13.5 & 4.5 \\
Gemma-9B/int4 & 516.4 & 98.3 &  93.6 &  84.3 & 7.5 & 6.8 & 4.7 & 14.4 & 2.4 & 30.2 & 14.0 & 4.5 \\
Ablit-4B/fp   &  18.5 & 19.1 & 184.5 & 106.3 & 2.7 & 0.8 & 7.6 &  6.7 & 1.2 &  7.7 &  1.0 & 1.1 \\
Ablit-4B/int8 &  17.0 & 17.0 & 180.7 & 103.8 & 2.6 & 0.9 & 7.4 &  6.6 & 1.2 &  7.8 &  0.9 & 1.1 \\
Ablit-4B/int4 &  15.2 & 13.5 & 188.7 & 111.2 & 2.1 & 0.5 & 6.0 &  4.8 & 1.8 &  8.1 &  0.8 & 0.8 \\
Qwen-1.7B/fp   & 219.6 & 39.8 & 45.9 & 38.4 & 6.4 & 2.0 & 8.1 & 8.8 & 4.5 & 7.4 & 0.9 & 2.1 \\
Qwen-1.7B/int8 & 211.2 & 37.2 & 46.5 & 38.6 & 6.1 & 1.7 & 8.1 & 8.8 & 3.7 & 6.1 & 0.8 & 1.8 \\
Qwen-1.7B/int4 & 194.8 & 41.7 & 50.7 & 38.5 & 5.7 & 1.7 & 7.0 & 8.0 & 3.3 & 4.7 & 0.5 & 1.7 \\
Qwen-4B/fp   & 442.4 & 72.4 &  93.5 & 102.1 & 5.1 & 1.6 & 6.1 & 5.8 & 2.0 & 6.0 & 2.8 & 2.1 \\
Qwen-4B/int8 & 438.2 & 70.0 &  90.4 &  98.8 & 4.7 & 1.4 & 6.0 & 5.7 & 2.0 & 6.1 & 2.7 & 2.1 \\
Qwen-4B/int4 & 364.4 & 54.9 & 105.4 & 105.8 & 4.4 & 1.2 & 4.9 & 3.9 & 1.3 & 6.4 & 2.0 & 1.8 \\
SafeRL-4B/fp   & 314.1 & 64.0 & 81.0 & 86.9 & 6.8 & 1.7 & 5.4 & 5.3 & 0.7 & 6.7 & 0.7 & 1.2 \\
SafeRL-4B/int8 & 313.0 & 61.6 & 79.2 & 85.6 & 6.3 & 1.6 & 5.4 & 5.2 & 0.7 & 6.7 & 0.7 & 1.2 \\
SafeRL-4B/int4 & 275.7 & 53.1 & 87.8 & 92.0 & 5.9 & 1.6 & 3.1 & 3.9 & 1.0 & 8.3 & 0.4 & 0.8 \\
Qwen-8B/fp   & 487.2 & 50.0 & 58.8 & 86.1 & 10.7 & 2.9 & 6.8 & 5.2 & 3.1 & 10.0 & 7.0 & 2.9 \\
Qwen-8B/int8 & 490.6 & 49.3 & 57.8 & 85.0 & 11.1 & 2.9 & 6.7 & 5.2 & 3.3 &  9.7 & 7.1 & 2.9 \\
Qwen-8B/int4 & 475.4 & 56.0 & 53.4 & 77.0 & 10.6 & 4.3 & 9.8 & 6.2 & 2.8 &  8.9 & 6.0 & 3.0 \\
\bottomrule
\end{tabular}
\end{table*}

\begin{table*}[p]
\centering\footnotesize
\setlength{\tabcolsep}{3.2pt}
\caption{Absolute per-prompt token weight (\textbf{DCG} metric), summed over layers, by
axis, model and weight regime. Same layout as Table~\ref{tab:abs-count}.}
\label{tab:abs-dcg}
\begin{tabular}{@{}l*{12}{c}@{}}
\toprule
& \multicolumn{2}{c}{Saf} & \multicolumn{2}{c}{Cmp} & \multicolumn{2}{c}{Eva}
& \multicolumn{2}{c}{Sof} & \multicolumn{2}{c}{Hed} & \multicolumn{2}{c}{Hrm} \\
\cmidrule(lr){2-3}\cmidrule(lr){4-5}\cmidrule(lr){6-7}\cmidrule(lr){8-9}\cmidrule(lr){10-11}\cmidrule(lr){12-13}
Model/$q$ & D & N & D & N & D & N & D & N & D & N & D & N \\
\midrule
Gemma-9B/fp   & 121.0 & 23.6 & 18.8 & 17.1 & 1.5 & 1.4 & 0.8 & 2.8 & 0.4 & 6.7 & 2.5 & 1.0 \\
Gemma-9B/int8 & 120.8 & 23.4 & 18.8 & 17.2 & 1.5 & 1.4 & 0.8 & 2.8 & 0.4 & 6.7 & 2.4 & 1.0 \\
Gemma-9B/int4 & 119.0 & 23.5 & 19.6 & 17.2 & 1.3 & 1.3 & 0.8 & 3.0 & 0.4 & 6.4 & 2.5 & 1.0 \\
Ablit-4B/fp   & 3.7 & 3.6 & 44.2 & 22.8 & 0.6 & 0.2 & 1.5 & 1.3 & 0.2 & 1.5 & 0.2 & 0.2 \\
Ablit-4B/int8 & 3.4 & 3.2 & 43.2 & 22.4 & 0.6 & 0.2 & 1.5 & 1.3 & 0.2 & 1.5 & 0.2 & 0.2 \\
Ablit-4B/int4 & 3.1 & 2.6 & 45.2 & 23.8 & 0.5 & 0.1 & 1.2 & 0.9 & 0.3 & 1.5 & 0.2 & 0.2 \\
Qwen-1.7B/fp   & 47.2 & 8.0 & 11.1 & 8.3 & 1.2 & 0.4 & 1.6 & 1.6 & 0.7 & 1.4 & 0.2 & 0.4 \\
Qwen-1.7B/int8 & 45.3 & 7.5 & 11.3 & 8.4 & 1.2 & 0.3 & 1.6 & 1.6 & 0.6 & 1.2 & 0.2 & 0.4 \\
Qwen-1.7B/int4 & 41.6 & 8.5 & 12.4 & 8.4 & 1.1 & 0.3 & 1.3 & 1.5 & 0.6 & 1.0 & 0.1 & 0.4 \\
Qwen-4B/fp   & 95.9 & 14.9 & 19.3 & 21.7 & 0.9 & 0.3 & 1.1 & 1.1 & 0.3 & 1.2 & 0.6 & 0.5 \\
Qwen-4B/int8 & 95.0 & 14.2 & 18.7 & 21.2 & 0.9 & 0.3 & 1.1 & 1.1 & 0.3 & 1.2 & 0.5 & 0.5 \\
Qwen-4B/int4 & 76.6 & 10.9 & 21.7 & 22.2 & 0.8 & 0.3 & 0.9 & 0.7 & 0.2 & 1.2 & 0.4 & 0.4 \\
SafeRL-4B/fp   & 66.5 & 12.6 & 16.3 & 17.9 & 1.5 & 0.4 & 1.0 & 1.0 & 0.1 & 1.3 & 0.1 & 0.2 \\
SafeRL-4B/int8 & 66.2 & 12.0 & 16.1 & 17.9 & 1.4 & 0.4 & 1.0 & 0.9 & 0.1 & 1.3 & 0.1 & 0.3 \\
SafeRL-4B/int4 & 57.3 & 10.5 & 18.1 & 18.8 & 1.3 & 0.3 & 0.5 & 0.7 & 0.2 & 1.5 & 0.1 & 0.2 \\
Qwen-8B/fp   & 107.7 & 10.2 & 11.4 & 17.5 & 2.0 & 0.7 & 1.3 & 1.0 & 0.5 & 1.8 & 1.4 & 0.7 \\
Qwen-8B/int8 & 107.9 & 10.0 & 11.1 & 17.2 & 2.1 & 0.7 & 1.3 & 1.0 & 0.6 & 1.8 & 1.4 & 0.6 \\
Qwen-8B/int4 & 105.7 & 11.4 & 10.1 & 15.4 & 2.0 & 0.9 & 1.9 & 1.2 & 0.5 & 1.7 & 1.2 & 0.7 \\
\bottomrule
\end{tabular}
\end{table*}

\begin{table*}[p]
\centering\footnotesize
\setlength{\tabcolsep}{4pt}
\caption{Mechanism localization at FP: number of important layers
$|\mathcal{I}_g|=|\{\ell:|\AUC_g(\ell)-0.5|>0.2\}|$ and the peak-$\AUC$ layer, reported as
``$|\mathcal{I}_g|$ / peak $\ell$''.}
\label{tab:important-layers}
\begin{tabular}{@{}lcccccc@{}}
\toprule
Model & Saf & Cmp & Eva & Sof & Hed & Hrm \\
\midrule
Gemma-9B  & 26/25 & 10/27 & 1/23 & 2/38 & 9/21 & 0/38 \\
Ablit-4B  &  2/4  & 17/28 & 0/15 & 0/29 & 0/25 & 0/29 \\
Qwen-1.7B & 12/17 &  1/12 & 7/24 & 0/26 & 3/17 & 0/15 \\
Qwen-4B   & 18/26 &  2/1  & 0/15 & 0/23 & 0/22 & 0/30 \\
SafeRL-4B & 18/21 &  1/0  & 6/28 & 4/27 & 1/34 & 0/26 \\
Qwen-8B   & 19/26 &  8/19 & 7/23 & 0/32 & 3/15 & 0/27 \\
\bottomrule
\end{tabular}
\end{table*}

\begin{table*}[p]
\centering\footnotesize
\setlength{\tabcolsep}{3.5pt}
\caption{Layer relocation under quantization: Jaccard $J$ of important-layer sets
$\mathcal{I}_g(\mathrm{FP})$ vs.\ $\mathcal{I}_g(q)$, with dropped
($\mathcal{I}_{\mathrm{FP}}\!\setminus\!\mathcal{I}_{\mathrm{int4}}$) and added layers under
INT4, for safety and compliance. Reported as an indicator only (\S\ref{sec:cross-quant}),
causal role is established separately.}
\label{tab:relocation}
\begin{tabular}{@{}llccll@{}}
\toprule
Model & Axis & $J$(i8) & $J$(i4) & Dropped (i4) & Added (i4) \\
\midrule
Gemma-9B  & safety     & 1.00 & 1.00 & --      & --        \\
Gemma-9B  & compliance & 0.90 & 0.90 & 20      & --        \\
Ablit-4B  & safety     & 0.00 & 0.00 & 2,4     & 10,12     \\
Ablit-4B  & compliance & 0.78 & 0.79 & 10,12   & 1,15      \\
Qwen-1.7B & safety     & 1.00 & 1.00 & --      & --        \\
Qwen-1.7B & compliance & 0.50 & 0.00 & 12      & --        \\
Qwen-4B   & safety     & 0.94 & 0.90 & --      & 12,14     \\
Qwen-4B   & compliance & 0.20 & 0.50 & 12      & --        \\
SafeRL-4B & safety     & 1.00 & 0.95 & --      & 10        \\
SafeRL-4B & compliance & 0.25 & 0.20 & --      & 4,19,21,34 \\
Qwen-8B   & safety     & 1.00 & 0.95 & 17      & --        \\
Qwen-8B   & compliance & 0.75 & 0.50 & 14,15   & 5,12,13,21 \\
\bottomrule
\end{tabular}
\end{table*}

\begin{table*}[p]
\centering\small
\setlength{\tabcolsep}{5pt}
\caption{Top-$k$ sensitivity of the safety axis at FP: best-layer $\AUC$ as a function of the counting depth $k$. A practical saturation point is reached at $k=100$, so this value is used for all model evaluations.}
\label{tab:topk}
\begin{tabular}{@{}lcccccc@{}}
\toprule
$k$ & Gemma-9B & Ablit-4B & Qwen-1.7B & Qwen-4B & SafeRL-4B & Qwen-8B \\
\midrule
5   & 0.921 & 0.559 & 0.826 & 0.918 & 0.910 & 0.955 \\
15  & 0.962 & 0.631 & 0.870 & 0.949 & 0.965 & 0.980 \\
30  & 0.974 & 0.760 & 0.889 & 0.969 & 0.977 & 0.978 \\
50  & 0.977 & 0.810 & 0.894 & 0.968 & 0.977 & 0.980 \\
100 & 0.983 & 0.770 & 0.917 & 0.973 & 0.984 & 0.983 \\
200 & 0.989 & 0.790 & 0.919 & 0.976 & 0.980 & 0.985 \\
\bottomrule
\end{tabular}
\end{table*}

\clearpage

\begin{figure*}[p]
\centering
\includegraphics[width=\textwidth]{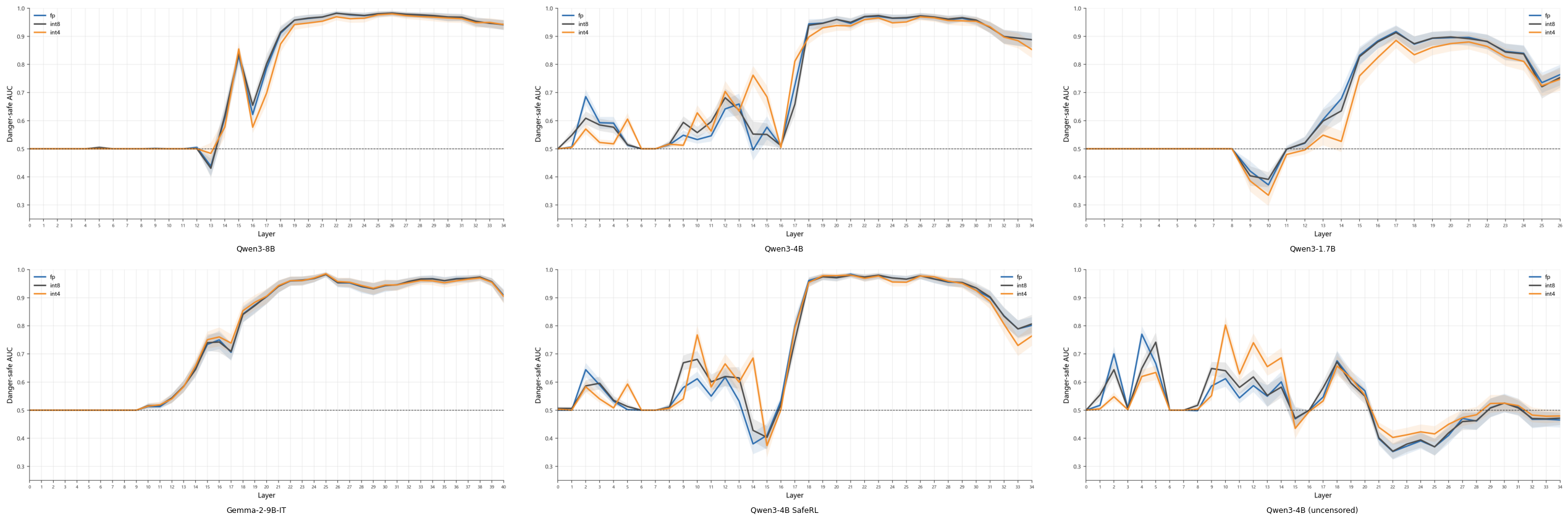}
\caption{Per-layer $\AUC_{\mathrm{safety}}(\ell)$ for the three weight-representation regimes. For
five models the profile has a characteristic shape: around $0.5$ in the early layers, then a sharp
transition and a plateau at $0.9$--$0.98$ in the middle and late part of the network. Ablit-4B (Qwen3-4B
uncensored) stands out with a qualitatively different profile: a brief rise in the mid-layer range with no
stable plateau, followed by a return to a level close to chance -- a visual confirmation of the weak
and unstable localization of the safety axis in this model, already noted from the numerical SafetyAUC$=0.593$.}
\label{fig:layer-auc-safety}
\end{figure*}

\begin{figure*}[p]
\centering
\includegraphics[width=\textwidth]{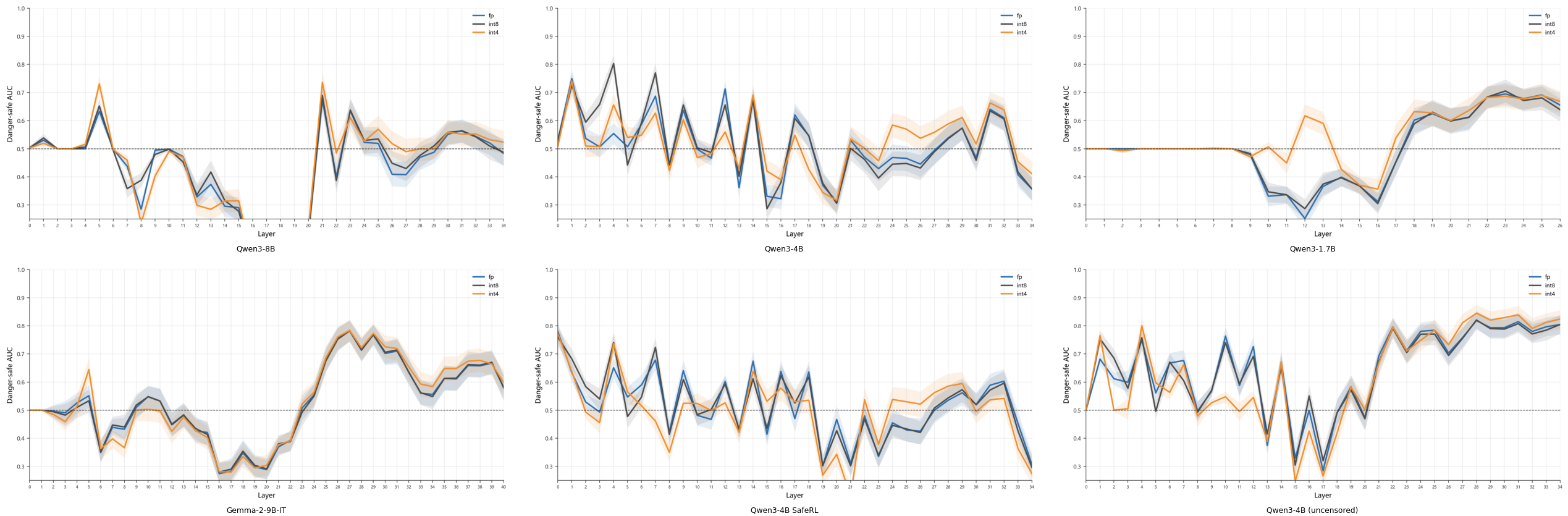}
\caption{Per-layer $\AUC_{\mathrm{compliance}}(\ell)$. SafeRL-4B already shows a pronounced AUC
spike at layer $0$ ($\approx 0.75$--$0.77$) -- no other model expresses the compliance signal this
strongly at the very first layer. This points to the nature of the model's training: it must not
simply refuse a harmful prompt, but is required to give some answer regardless, and this readiness
to answer is already visible in the representation of the prompt before any substantive
processing takes place.}
\label{fig:layer-auc-compliance}
\end{figure*}

\begin{figure*}[p]
\centering
\includegraphics[width=\textwidth]{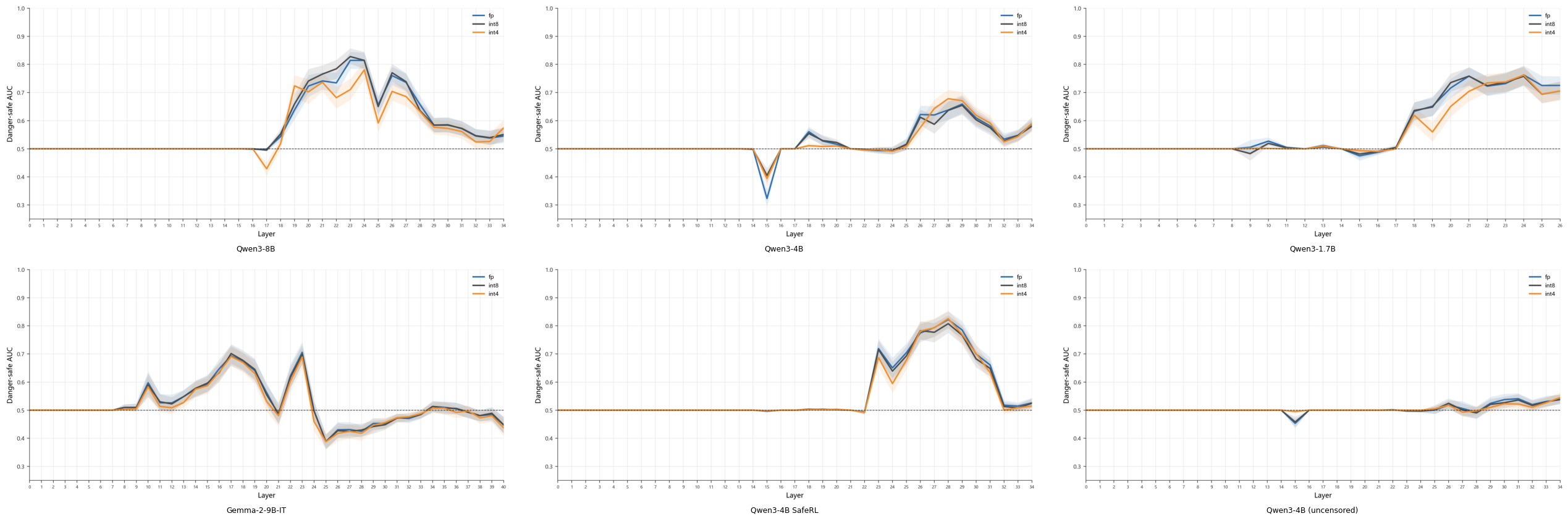}
\caption{Per-layer $\AUC_{\mathrm{evasion}}(\ell)$. For most models the axis is localized in a
narrow band of mid-to-late layers ($\AUC$ up to $0.7$--$0.85$), for Ablit-4B the profile stays
close to $0.5$ across the entire range of layers.}
\label{fig:layer-auc-evasion}
\end{figure*}

\begin{figure*}[p]
\centering
\includegraphics[width=\textwidth]{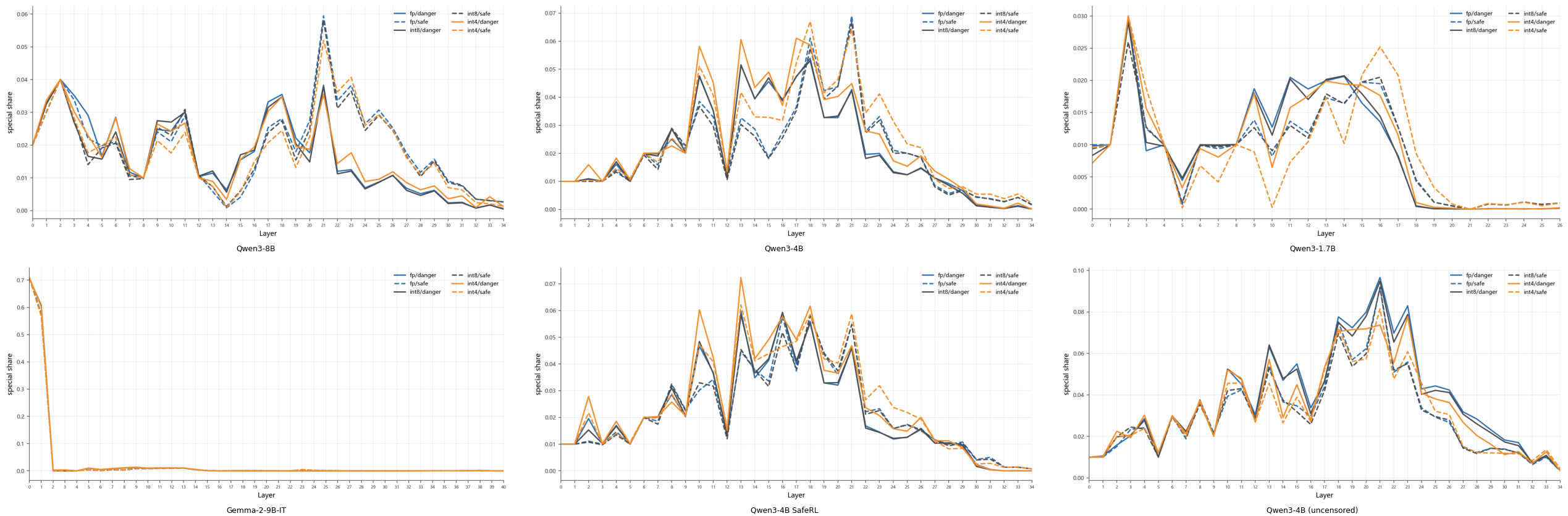}
\caption{Share of special/technical tokens in the top-$k$ readout by layer, for four
regime/promptset combinations. For Gemma-2-9B-IT almost the entire mass of special tokens is
concentrated in the first and second layer ($\approx 0.6$--$0.7$), after which the share falls to
near zero and stays there across the rest of the range. For models in the Qwen3 family, by
contrast, the special-token share fluctuates between $0.01$ and $0.07$ across nearly all layers. This
is a notable architectural difference between the model families.}
\label{fig:special-share}
\end{figure*}

\begin{figure*}[p]
\centering
\includegraphics[width=\textwidth]{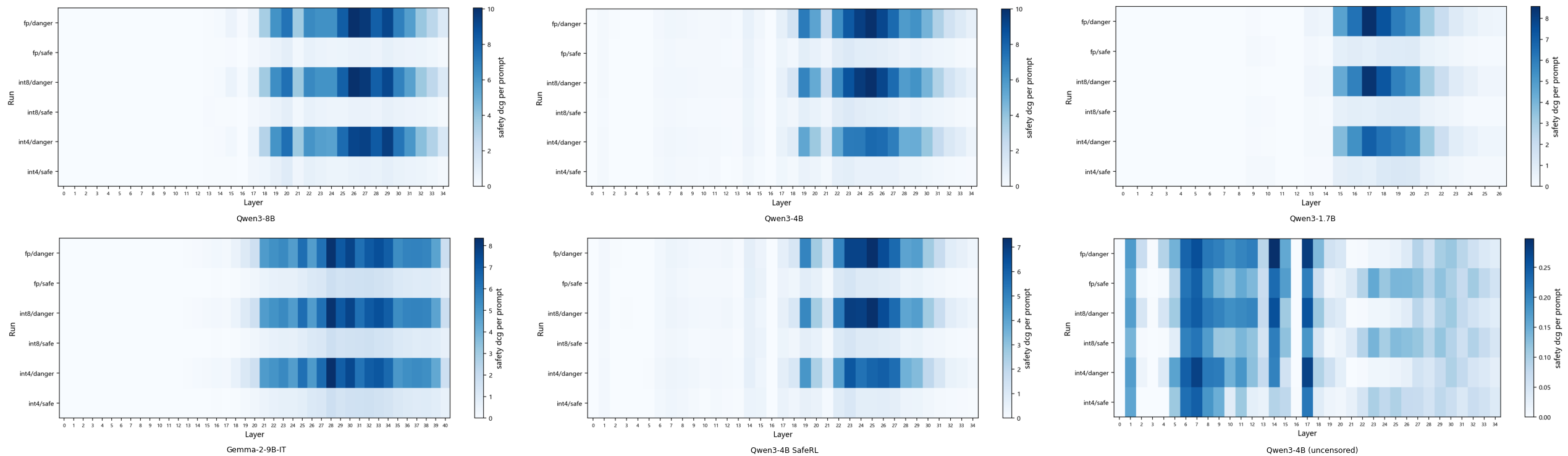}
\caption{Heatmap of safety DCG by layer and regime (danger/safe $\times$ $\fp$/INT8/INT4).
Five models show sharp, narrowly localized dark bands only in the danger rows, in the
mid-to-late range of layers. For Ablit-4B the intensity scale is an order of magnitude lower, and the dark
regions are spread over a far larger number of layers while barely telling danger from safe --
the visual counterpart of a low and unstable SafetyAUC.}
\label{fig:safety-dcg-heatmap}
\end{figure*}

\begin{figure*}[p]
\centering
\includegraphics[width=\textwidth]{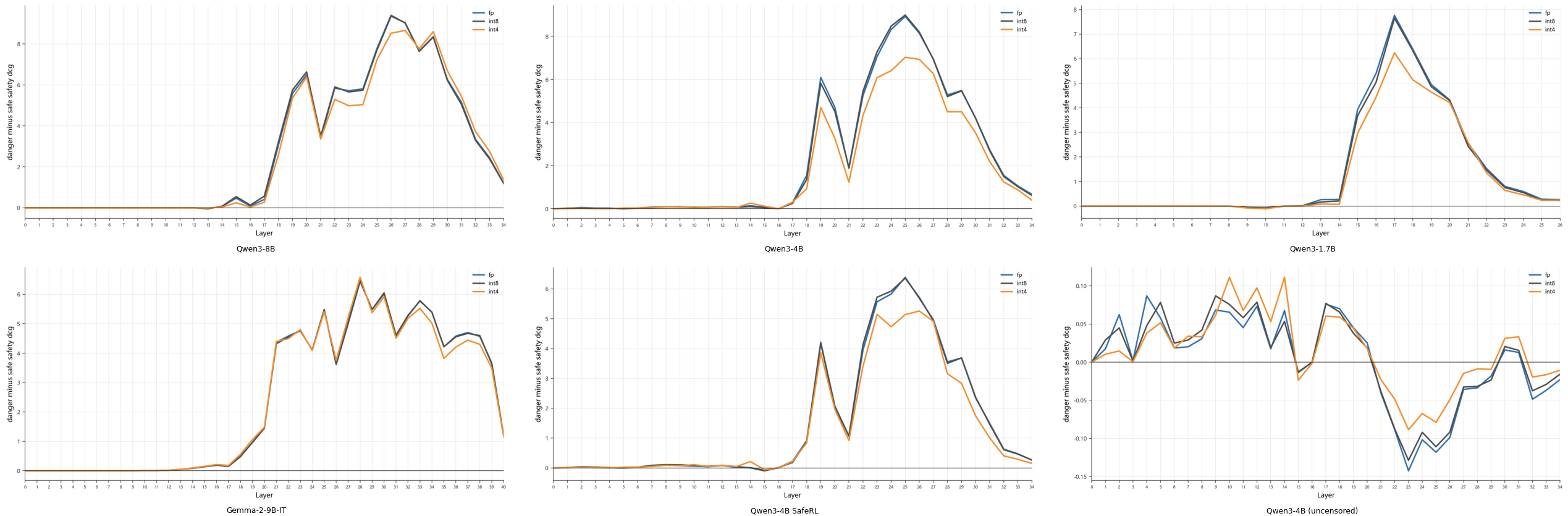}
\caption{Difference (danger minus safe) in safety DCG by layer. Five models show a clean
dome-shaped profile with an amplitude of $6$--$9$ DCG units. For Ablit-4B the amplitude is two orders
of magnitude smaller ($-0.15$ to $0.11$) and turns negative over part of the layer range
(roughly 22--29), meaning the safe set carries more safety vocabulary than the danger set in these
layers -- an inversion found in no other model.}
\label{fig:safety-dcg-delta}
\end{figure*}

\begin{figure*}[p]
\centering
\includegraphics[width=\textwidth]{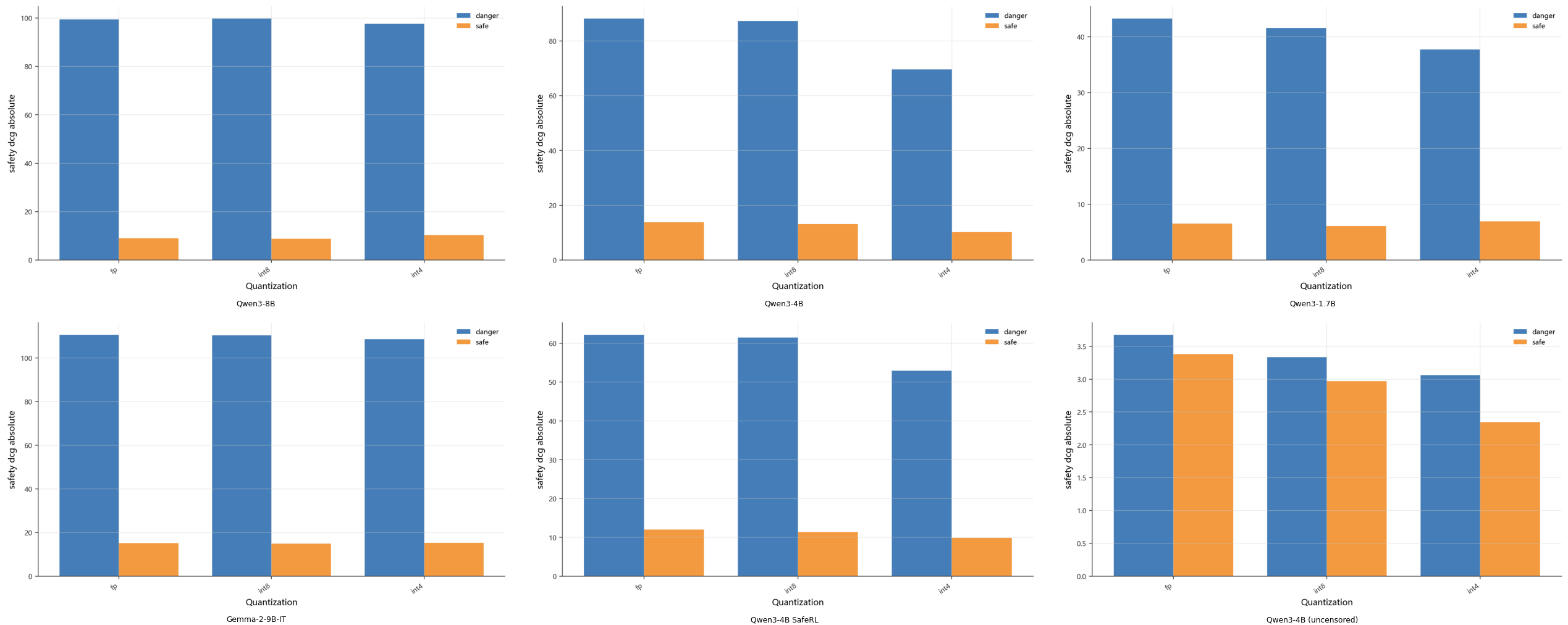}
\caption{Absolute safety DCG mass on danger and safe, by model and quantization regime
(numerical values in Table~\ref{tab:abs-dcg} of the Appendix). A basic descriptive plot for
comparison against the rank metrics of the main text.}
\label{fig:safety-dcg-absolute}
\end{figure*}

\begin{figure*}[p]
\centering
\includegraphics[width=\textwidth]{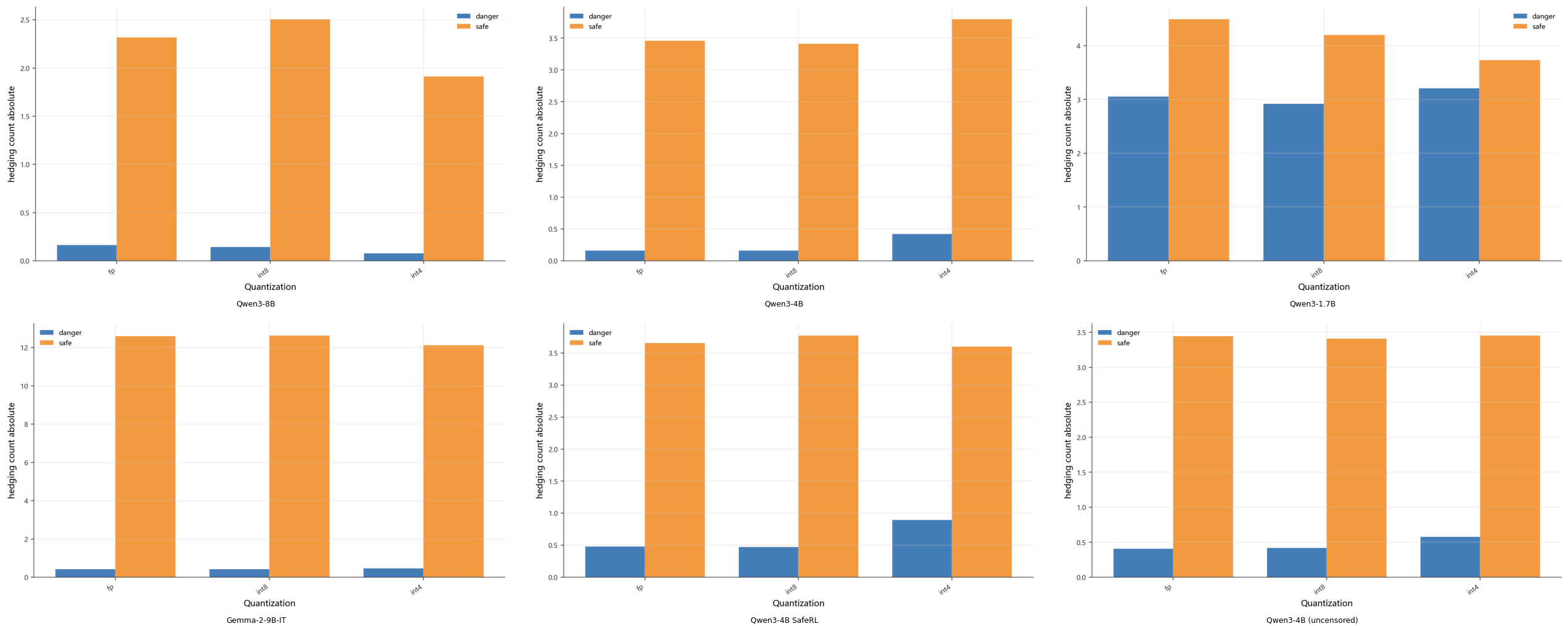}
\caption{Absolute hedging count mass on danger and safe. For five models, hedging is almost
entirely concentrated on the safe set, and is close to zero on danger. Qwen3-1.7B is the sole
exception: its value on danger ($\approx 3.0$--$3.3$) is comparable in order of magnitude to its
value on safe ($\approx 3.7$--$4.5$). For a very small model, a rise in hedging on dangerous
prompts may be an indirect sign of compliance-like behavior -- an evasive, cautious, yet
non-refusing response -- rather than mere noise from a small model.}
\label{fig:hedging-count-absolute}
\end{figure*}

\begin{figure*}[p]
\centering
\includegraphics[width=\textwidth]{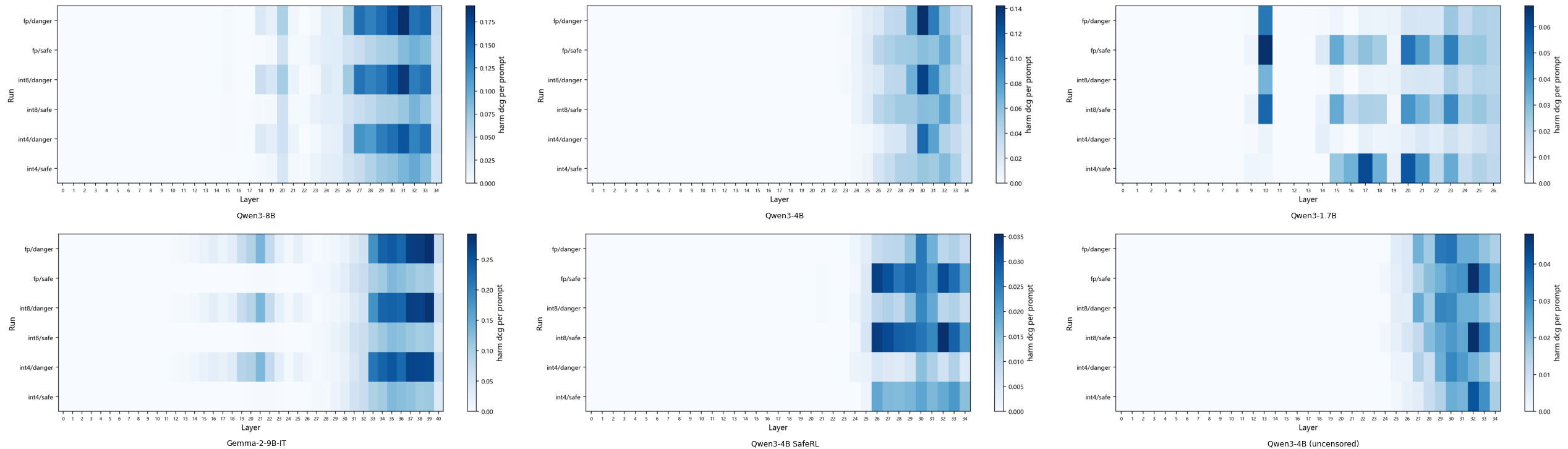}
\caption{Heatmap of harm DCG by layer and regime. For SafeRL-4B the fp/safe and int8/safe
rows show noticeably darker and wider bands than the corresponding danger rows of the same
model -- for the other models the relation runs the other way (danger darker than safe). This
agrees with the observation in \S\ref{sec:results}: SafeRL-4B is trained to explicitly name the
harmful domain of a prompt and to answer with an extended, safe explanation rather than the short
refusal that is sufficient for the other models and does not call for activating harm vocabulary on
the safe set.}
\label{fig:harm-dcg-heatmap}
\end{figure*}

\begin{figure*}[p]
\centering
\includegraphics[width=\textwidth]{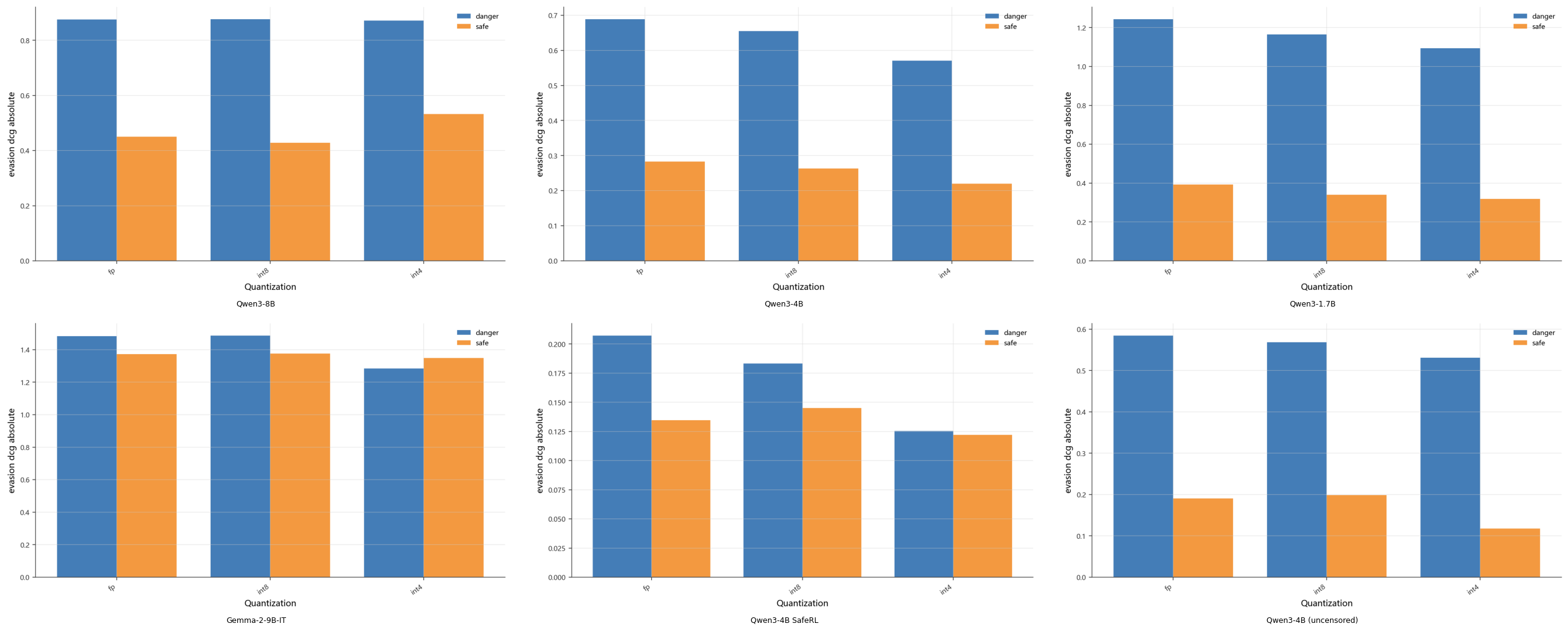}
\caption{Absolute evasion DCG mass on danger and safe. For Gemma-2-9B-IT, danger consistently
exceeds safe at $\fp$ and INT8, but the relation flips under INT4 (safe becomes higher than
danger) -- the only such case among the six models. This is consistent with Gemma being trained
more strongly than the others to resist attempts at evasion and consequently behaving more
strictly: after quantization, its evasion response to harmless but lexically alarming safe prompts
starts to outweigh its response to genuinely dangerous ones.}
\label{fig:evasion-dcg-absolute}
\end{figure*}

\begin{figure*}[p]
\centering
\includegraphics[width=\textwidth]{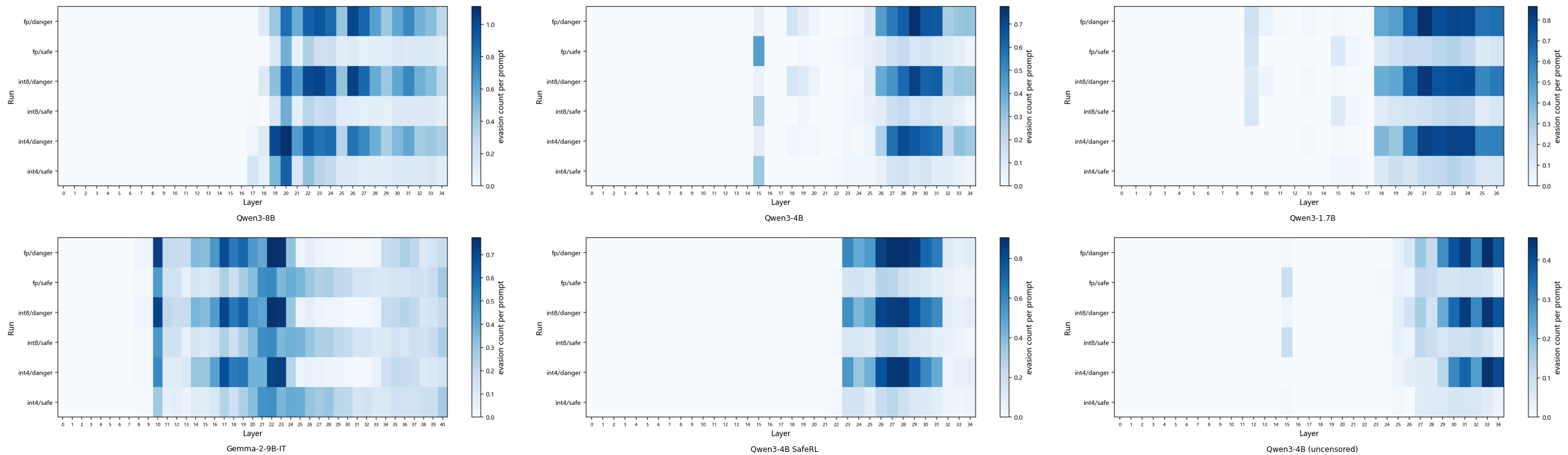}
\caption{Heatmap of evasion count by layer and regime. For Gemma-2-9B-IT the signal appears
noticeably earlier in network depth (around layer 9--10) than for models of the Qwen3 family
(typically after layer 17-20) -- further confirmation that Gemma's resistance to evasion attempts
is built into earlier, not only later, stages of processing the prompt.}
\label{fig:evasion-count-heatmap}
\end{figure*}

\begin{figure*}[p]
\centering
\includegraphics[width=\textwidth]{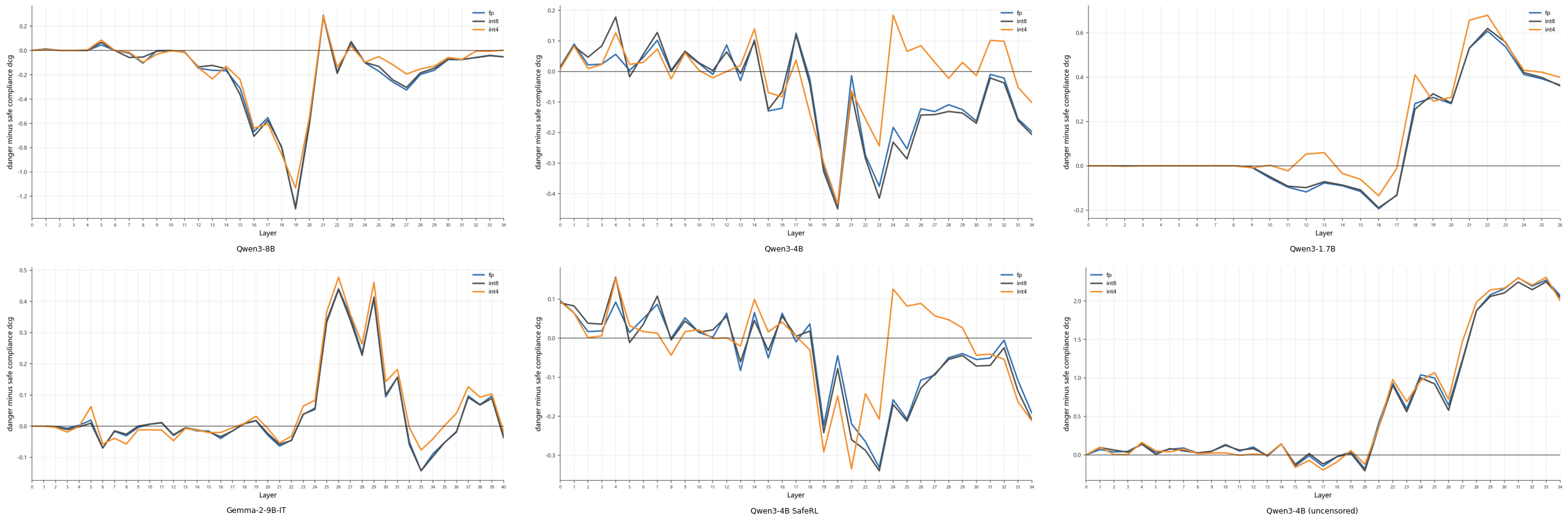}
\caption{Difference (danger minus safe) in compliance DCG by layer. Unlike safety
(fig.~\ref{fig:safety-dcg-delta}), the profiles here are noticeably narrower, noisier, and do not
form a single dome of consistent sign across all models -- the compliance axis is less well
localized and less stable across layers, consistent with the smaller number of important layers
$|\mathcal{I}_{\mathrm{compliance}}|$ in Table~\ref{tab:important-layers}.}
\label{fig:compliance-dcg-delta}
\end{figure*}

\end{document}